# Magnetic Untwisting in Solar Jets that Go into the Outer Corona in Polar Coronal Holes

Short title: **Magnetic Untwisting in Solar Jets**

Ronald L. Moore[1,2], Alphonse C. Sterling[1], and David A. Falconer[1,2]
[1] Heliophysics and Planetary Science Office, ZP13, Marshall Space Flight Center, Huntsville, AL 35812, USA; ron.moore@nasa.gov
[2] Center for Space Plasma and Aeronomic Research, University of Alabama in Huntsville, AL 35899, USA

ABSTRACT

We study 14 large solar jets observed in polar coronal holes. In EUV movies from *SDO*/AIA, each jet appears similar to most X-ray jets and EUV jets that erupt in coronal holes, but each is exceptional in that it goes higher than most, so high that it is observed in the outer corona beyond 2.2 $R_{Sun}$ in images from the *SOHO*/LASCO/C2 coronagraph. From AIA He II 304 Å movies and LASCO/C2 running-difference images of these high-reaching jets, we find: (1) the front of the jet transits the corona below 2.2 $R_{Sun}$ at a speed typically several times the sound speed; (2) each jet displays an exceptionally large amount of spin as it erupts; (3) in the outer corona, most of the jets display measureable swaying and bending of a few degrees in amplitude; in three jets the swaying is discernibly oscillatory with a period of order 1 hour. These characteristics suggest that the driver in these jets is a magnetic-untwisting wave that is basically a large-amplitude (i.e., non-linear) torsional Alfven wave that is put into the reconnected open field in the jet by interchange reconnection as the jet erupts. From the measured spinning and swaying we estimate that the magnetic-untwisting wave loses most of its energy in the inner corona below 2.2 $R_{Sun}$. We point out that the torsional waves observed in Type-II spicules might dissipate in the corona in the same way as the magnetic-untwisting waves in our big jets and thereby power much of the coronal heating in coronal holes.

## 1. INTRODUCTION

*SOHO* (*Solar & Heliospheric Observatory*) has nearly continuously observed the Sun since 1996, when the sunspot cycle was passing through its minimum between Cycle 22 and Cycle 23. The C2 coronagraph, of the Large Angle Spectroscopic Coronagraph (LASCO) instrument on *SOHO*, images in white light the outer corona between 2 $R_{Sun}$ and 6 $R_{Sun}$ at a nominal cadence of 12 minutes (Brueckner et al 1995). The Extreme-Ultraviolet Telescope (EIT) on *SOHO*, until its operation was curtailed after the *Solar Dynamics Observatory* (*SDO*) (Pesnell et al 2012) became operational in 2010, imaged the inner corona in the 195 Å emission from Fe XII at a nominal cadence of 12 minutes, and at slower cadence in three other EUV bands: one for the 304 Å emission from He II, one for the 171 Å emission from Fe IX-X, and one for the 284 Å emission from Fe XV (Delaboudinere et al 1995). In the LASCO/C2 movies spanning ten months (1997 April 2 – 1998 February 8) soon after sunspot minimum, Wang et al (1998) found in the north and south polar coronal holes 27 white-light jets that came from EUV jets observed at the base of the coronal hole in the EIT 195 Å movies. In the LASCO/C2 images, these jets had angular widths of a few degrees. For each jet, from the EIT 195 Å images and/or the C2 white-light images, Wang et al (1998) measured three speeds, only one of which is dealt with in this paper. That one we call the jet-front transit speed, the constant speed that would move the tip of a jet from its point in the low corona in an EIT 195 Å image to its point in the outer corona in the first C2 image of the jet in the time elapsed between the two images. The 27 jets had jet-front transit speeds that ranged from 410 km $s^{-1}$ to 1090 km $s^{-1}$, and averaged 630 km $s^{-1}$. The EIT 195 Å images showed obvious helical twist in one of the 27 jets. Wang et al (1998) pointed out that the EUV jets that generated the jets seen in the outer corona by LASCO/C2 were similar to X-ray jets seen by the Soft X-ray Telescope (SXT) on *Yohkoh* and were plausibly produced in the manner proposed by Shibata et al (1992) for X-ray jets, by reconnection of closed magnetic field in the base of the jet with the ambient open magnetic field of the coronal hole.

Moore et al (2010) examined the magnetic structure displayed by X-ray jets in coronal X-ray movies of the polar coronal holes taken by the X-Ray Telescope (XRT) on *Hinode* (Golub et al 2007). The movies they studied were taken in 2008, when the sunspot cycle was again passing through minimum. Only jets that were large enough to clearly display substructure in the XRT images were selected for study, jets having base widths of $\sim$ 20,000 km or more. They found that X-ray jets of this size occurred at rate of about 1 jet per hour per polar coronal hole, confirming the rate of about 30 jets per day per polar coronal hole found by Savcheva et al (2007) from XRT movies of the polar coronal holes. The EUV-jet sources of the white-light polar jets found by Wang et al (1998) were also of this size. Many polar X-ray jets of this size in XRT images are also visible in coronal EUV images such as those from EIT (Raouafi et al 2008; Moore et al 2010, 2013; Pucci et al 2013). Therefore, many if not all of the EUV jets of Wang et al (1998) plausibly would have been seen as X-ray jets in XRT movies, had XRT existed then and had taken coronal X-ray movies of Wang's EUV jets. During 1996, when the polar coronal jets studied by Wang et al (1998) were observed, the polar coronal holes were noticeably larger in area than during 2008, when the polar X-ray jets studied by Moore et al (2010) were observed. Even so, Wang et al (1998) found the rate of occurrence of the white-light jets detected in polar coronal holes by LASCO/C2 was only 1-2 jets per day per polar coronal hole, 15 – 30 times less than the rate of occurrence, in smaller polar coronal holes, of X-ray jets of the size of the EUV jets that generated the white-light jets. In agreement with the recent study of polar jets by Yu et al (2014), these observed rates of occurrence indicate that only a tiny minority of EUV jets in the polar coronal holes, no more than a few percent of the big ones, propagate into the outer corona beyond 2 $R_{Sun}$ with enough plasma to be seen in LASCO/C2 images. Conversely, the rest, the other big ones and the vastly more numerous smaller ones (the smaller EUV macrospicules and the still smaller EUV spicules), either do not go that high or are too faint there to be individually detected by LASCO/C2.

From the observed structure of X-ray jets as they erupt in the polar coronal holes, Moore et al (2010) concluded that coronal jets (X-ray jets and EUV and H$\alpha$ macrospicules) in coronal holes are of two different kinds, which they named standard jets (denoting that these jets appear to fit the widely-accepted Shibata et al (1992) model for coronal jets) and blowout jets (denoting the way these jets erupt). Moore et al (2010) inferred/proposed from their study of XRT X-ray jets that in either kind of jet, there is a closed-arch bipolar magnetic field in the base of the jet, a closed magnetic arch embedded in the open field of the coronal hole, and in the eruption of the jet interchange reconnection occurs on the outside of the closed field more or less in the manner proposed by Shibata et al (1992), external reconnection of base-arch closed field with oppositely-directed ambient open field. The essential difference between the two kinds of jets is that the field far inside the base arch is drastically more active in blowout jets than in standard jets. A standard jet is entirely produced by a burst of interchange reconnection of the outside of the base arch, during which the interior of the base arch remains closed. In contrast, a blowout jet is produced by blowout eruption of the interior field of the base arch, that is, by eruption of the base arch as in the blowout of a closed-arcade field in the production of a major coronal mass ejection (CME) (For a well-observed EUV blowout jet that definitely shows its mini-CME nature, see Hong et al 2011). In a blowout jet, the base-arch interior field presumably has enough free energy (enough shear and twist) to blowout, whereas in a standard jet the interior field presumably does not have enough free energy to blowout. Blowout of the base arch drives more widespread interchange reconnection than occurs in a standard jet, producing more strands of reconnected open field with jet outflow on them, and at the same time ejects plasma carried inside the blowing-out closed field. This makes the spray of ejected plasma (the spire of the jet) wider and more complex in blowout jets than in standard jets. Also, during the eruptive growth phase, because there is more reconnection in the interior of the base arch in a blowout jet than in a standard jet, the interior of the base arch is heated more and is markedly brighter in X-ray emission than in a standard jet.

The magnetic field's different form and action that Moore et al (2010) envisioned for standard jets and blowout jets implies that in the eruptive growth phase of coronal-hole X-ray jets having base widths of ~ 20,000 km or more, most blowout jets should be much more visible than most standard jets in He II 304 Å movies, because (1) most blowout jets should have lots of plasma at cool-transition-region temperatures ($T \sim 10^5$ K) carried in field that blows out from low in the base arch, as in an ejective filament-eruption flare, and (2) the interchange reconnection in most standard jets of this size should occur in the low corona above the transition region and hence should eject little or no plasma having sub-coronal temperature ($T < 10^6$ K). For four of the more than 50 XRT X-ray jets studied by Moore et al (2010), two standard jets and two blowout jets, the few-minute eruptive growth phase was caught in a full-disk He II 304 Å snapshot from *STEREO*/EUVI (*Solar Terrestrial Relations Observatory*/Extreme UltraViolet Imager) (Howard et al 2008), which images have 1.6 arcsec pixels and a cadence of 10 min or slower. These He II 304 Å snapshots provided a preliminary positive test of the proposed dichotomy scenario for coronal jets: the two blowout X-ray jets had an obvious He II 304 Å cool component in their ejecta and the two standard X-ray jets had none.

Moore et al (2013) used the full-disk He II 304 Å movies from the Atmospheric Imaging Assembly (AIA) (Lemen et al 2012) on SDO to study the cool (T ~ $10^5$ K) component of X-ray jets observed in polar coronal holes by XRT. At the outset of this study, the goal was to further test the standard-jet/blowout-jet dichotomy of polar X-ray jets proposed in Moore et al (2010). In addition to verifying the dichotomy of polar X-ray jets, the AIA 304 Å movies, via their high cadence (12 s) and high spatial resolution (0.6 arcsecond pixels), revealed that most polar X-ray jets spin as they erupt. Moore et al (2013) examined 54 X-ray jets that were found in polar coronal holes in XRT movies sporadically taken during the first year of continuous operation of AIA (2010 May through 2011 April), and that were big enough and bright enough in the XRT images to be judged to be a standard jet or a blowout jet. From the X-ray movies, 19 of the 54 jets appeared to be standard jets, 32 appeared to be blowout jets, and 3 were ambiguous. As was anticipated in Moore et al (2010), the standard-jet/blowout-jet dichotomy of X-ray jets found from XRT movies was confirmed by the visibility of the 54 X-ray jets in the AIA 304 Å movies: the ejecta of nearly all of the blowout jets (29 of 32) had a cool component that was visible in the 304 Å movies, and nearly all of the standard jets (16 of 19) had no cool component, no ejecta visible in the 304 Å movies. (None of the 3 ambiguous X-ray jets had any visible ejecta in the 304 Å movies.) In agreement with the different magnetic genesis envisioned in Moore et al (2010) for standard jets and blowout jets: in all 29 blowout X-ray jets in which the ejecta had a He II 304 Å cool component, the cool component of the spire displayed obvious lateral expansion as it erupted; whereas in all 3 standard X-ray jets in which the ejecta had a He II 304 Å cool component, the cool component displayed no appreciable lateral expansion. The 304 Å movies also revealed that there is discernible axial rotation, spinning motion about the axis of the spire, in the eruptive growth phase in most (perhaps all) of the larger X-ray jets in the polar coronal holes: of the 32 X-ray jets that could be seen in the 304 Å movies, 29, including all 3 of the standard jets that displayed a cool component, displayed measurable axial rotation. (Three of these 32 were blowout jets in which the 304 Å cool component was too sparse and brief for any axial rotation to be discerned.) From the 304 Å movies, Moore et al (2013) measured the speed of the axial rotation near the outer edge of the spire in one standard jet and two blowout jets and found it to be of order 60 km $s^{-1}$. There are many earlier published observations of untwisting spiral structure and untwisting speeds of this order in the spires of large coronal jets (Pike & Mason 1998; Patsourakos et al 2008; Kamio et al 2010; Raouafi et al 2010; Sterling et al 2010a; Curdt et al 2012**;** Morton et al 2012).

For the larger X-ray jets (base width ~ 20,000 km or more), when the polar coronal holes are biggest, during the minimum phase of the sunspot cycle, there is on average ~ 1 X-ray jet occurring at any instant in both polar coronal holes together (Savcheva et al 2007; Moore et al 2010). The upper chromosphere in quiet regions and coronal holes is a forest of chromospheric-temperature jets that are much smaller and vastly more numerous than the X-ray jets in coronal holes. Globally, in the quiet regions and coronal holes all over the Sun, there are always ~ $10^6$ chromospheric jets occurring at any instant, ~ $10^2$ jets per supergranule (Moore et al 2011). These chromospheric jets are the so-called Type-II spicules discovered by the unprecedented steady 0.2 arcsec resolution and ~ 10 sec cadence of the Ca II H movies from the Solar Optical Telescope (SOT) on *Hinode* (e.g., De Pontieu et al 2007a). They have widths of ~ 300 km, lifetimes of 1-3 minutes, outflow speeds of 50-100 km $s^{-1}$, oscillatory lateral swaying motion having periods of a few minutes and speeds of ~ 20 km $s^{-1}$, and axial rotation (spin) speeds of 25-30 km $s^{-1}$ (De Pontieu et al 2007b, 2012; Pereira et al 2012).

From a SOT Ca II movie of the chromosphere at the limb in a polar coronal hole, Sterling et al (2010b) found that Type-II spicules display a dichotomy comparable to the standard-jet/blowout-jet dichotomy of X-ray jets in that many Type-II spicules are single strands that do not expand laterally and many others display multiple strands that laterally expand apart. From the lateral-expansion dichotomy of Type-II spicules and from the discovery by the *Hinode*/SOT SpectroPolarimeter (SP) that everywhere in quiet regions and coronal holes there is a granule-size (diameter ~ 1000 km) emerging magnetic bipole (magnetic arch) in about 1 out of every 10 granules (Ishikawa et al 2008), Moore et al (2011) proposed that Type-II spicules are tiny counterparts of X-ray jets, i.e., that Type-II spicules erupt from granule-size emerging bipoles in the same ways that X-ray jets erupt from larger closed bipolar magnetic fields. This idea is also in accord with X-ray jets and Type-II spicules both typically having definite axial rotation (Moore et al 2013). Moore et al (2011) supposed that the transverse motions observed in Type-II spicules are basically Alfven waves that are generated by the burst of interchange reconnection in the jet-eruption process. They estimated that, averaged over areas of the Sun larger than a supergranule, the energy flux carried into the corona by the Alfven waves co-generated with Type-II spicules is ~ 3 x $10^5$ erg $cm^{-2}$ $s^{-1}$. This is of the order of the energy flux needed to power the heating of the corona in coronal holes (Withbroe 1988; Dobrzycka et al 2002; McIntosh et al 2011).

Based on the characteristics of solar jets summarized above, in interpreting the jet observations of the present paper, we assume: (1) that there is substantial twist in closed magnetic field in the base of practically all solar jets, (2) that, as proposed by Shibata & Uchida (1986) and by Canfield et al (1996), the interchange reconnection in the production of the jet transfers twist from the closed field to the reconnected open field, and (3) that the spin observed in X-ray and EUV jets as they erupt in coronal holes is that of the resulting wave of magnetic untwisting that propagates the transferred twist out along the open field. We take this magnetic-untwisting wave to be basically a torsional Alfven wave. The similarity of Type-II spicules to the much larger X-ray and EUV jets in coronal holes raises the possiblity that the transverse motions observed in Type-II spicules are from similar magnetic-untwisting waves that are generated by twist transfer by interchange reconnection as in the larger jets.

As the estimates of Moore et al (2011) show, the torsional magnetic waves accompanying Type-II spicules plausibly carry enough energy to power the coronal heating in coronal holes. Even if these waves have enough energy for the heating, whether they actually do power the heating remains an open question. An alternative possibility is that these waves do not dissipate much in the corona but carry most of their energy on out into the solar wind. In this paper, we investigate this question by analyzing the untwisting and swaying motions in high-reaching jets observed in the polar coronal holes, jets that were observed during their birth at the base of the corona, by AIA, and then in the outer corona, by LASCO/C2. Our goal is to discern whether the magnetic-untwisting waves in these big jets lose much of their energy in the inner corona below 2 $R_{Sun}$. On the basis of the similarity of solar jets of all sizes, it is plausible that the torsional waves in Type-II spicules deposit their energy in the inner corona to a similar degree and by the same process as in the big jets, whatever that process may be. Thus, our analysis of observed transverse motions in big jets that go into the outer corona could either strengthen or weaken the case for the corona in coronal holes being heated by the magnetic-untwisting waves that are co-generated with Type-II spicules.

## 2. OBSERVED JETS

### *2.1. Our Set of Polar Jets Observed by Both SDO/AIA and SOHO/LASCO/C2*

Because the SOHO LASCO CME Catalog (Yashiro et al 2004; Gopalswamy et at 2009) is available on-line and includes many polar jets observed in the outer corona by LASCO/C2, we used it to find the high-reaching polar jets studied in this paper. We found our jets by viewing, along with concurrent AIA EUV movies, the CME Catalog's LASCO/C2 running-difference movies of very narrow events, events for which the heliocentric angular width given in the Catalog was less than 10°. Each event that we identified to be a jet and selected for study (1) stemmed from a polar coronal hole, (2) was clearly seen in the C2 movie to be a jet, similar in appearance to the jets found by Wang et al (1998), and (3) evidently came from an EUV jet observed at the base of the coronal hole by AIA (the EUV jet had the appropriate position and the jet entered the C2 field of view (reached above 2.2 $R_{Sun}$) in less than about an hour after the start of the EUV jet). In the 19 months from 2010 August through 2012 March, during which time the polar coronal holes were shrinking in the rising phase of the sunspot cycle, we found in the LASCO/C2 running-difference movies 18 polar jets for which the source EUV jet was observed in the AIA movies. We studied the entire life of each jet, at a cadence of 12 s, in the movies from two AIA EUV channels: the AIA 304 Å channel, which images He II 304 Å emission from plasma at temperatures around 5 x $10^4$ K (in the $10^4$-$10^5$ K temperature range of the cooler transition region), and the AIA 193 Å channel, which images Fe XII 193 Å emission from coronal plasma at temperatures around 1.5 x $10^6$ K (Lemen et al 2012). From the observed character of each jet in these movies, 2 of these 18 jets were evidently standard jets and the rest were obviously blowout jets. The two standard jets and one of the blowout jets showed no ejecta in the He II 304 Å movies, and the He II 304 Å ejecta in another blowout jet was too sparse to show measurable axial rotation. Each of the other 14 blowout jets had more copious He II 304 Å ejecta, and this component of the jet showed measureable axial rotation. Of the 18 jets, we selected only these 14 for further study, because we were especially interested in the axial rotation in jets, and the axial rotation can be measured more accurately from the 304 Å movies than from the 193 Å movies. (Both movies have the same cadence (12 s) and the same resolution (0.6 arcsecond pixels), but structures in the jet (strands and clumps of plasma) can be seen more clearly and tracked with greater fidelity in the 304 Å movie than in the 193 Å movie. Consequently, the jet's axial rotation, the rotation of the structures about the axis of the jet as the jet erupts, is better measured from the 304 Å movie than from the 193 Å movie.) The date, time, and azimuthal position of each of these 14 jets are given in Table 1.

In Table 1, the date of each jet is in the first column, and the time of the start of the jet's eruption in the 304 Å movie is in the second column. The entries in the third, fourth, and fifth columns are from the *SOHO* LASCO CME Catalog. They are: the time of the C2 running-difference image in which the jet first appeared, the time of the last C2 running-difference image in which the jet had not yet faded into the background, and the position angle of the jet counterclockwise from solar north in the C2 images. The fifth column shows that 5 of the 14 jets happened in the northern polar coronal hole and 9 in the southern polar coronal hole, which is consistent with the northern coronal hole being considerably smaller than the southern one during the time interval of the search. The third and fourth columns together give the visible duration of each jet in the C2 running-difference movie. The duration ranged from 3 to 6 frames (24 to 60 minutes) during which time the fronts of the faster and longer-lasting jets reached beyond 5 $R_{Sun}$ (see Table 2), nearly to the 6 $R_{Sun}$ outer edge of the C2 field of view, as did many of the jets of Wang et al (1998). The second and third columns together give the transit time of each jet, the time between the start of the jet at the base of the corona (i.e., when first seen in the 304 Å movie) and the time of the jet's first appearance in C2 images of the outer corona. The transit time ranged from 25 to 92 minutes in our 14 jets (see Table 2). For the 27 polar jets of Wang et al (1998), the transit time ranged from 17 to 60 minutes. Thus, the data in Tables 1 and 2 establish the similarity of the jets in our set to those in the twice-larger set of polar jets studied by Wang et al (1998).

**Table 1**
Polar Jets Observed by Both SDO/AIA and SOHO/LASCO/C2

| Date | Times (UT) | | | Position Angle[a] in C2 Images (degrees) |
|---|---|---|---|---|
| | Eruption Start in AIA 304 Movie | First C2 Image | Last C2 Image | |
| 2010 Aug 11 | 09:53 | 10:36 | 11:24 | 11 |
| 2010 Aug 11 | 18:50 | 19:24 | 20:12 | 336 |
| 2010 Aug 19 | 20:43 | 21:24 | 22:12 | 12 |
| 2010 Sep 29 | 21:35 | 22:12 | 23:06 | 191 |
| 2010 Nov 8 | 21:45 | 22:36 | 23:17 | 349 |
| 2010 Nov 14 | 14:04 | 15:36 | 16:12 | 197 |
| 2010 Dec 4 | 21:42 | 22:24 | 23:06 | 165 |
| 2011 Jan 14 | 16:47 | 17:36 | 18:12 | 188 |
| 2011 Apr 9 | 05:35 | 06:00 | 06:48 | 188 |
| 2011 May 3 | 04:10 | 04:48 | 05:48 | 343 |
| 2011 Jun 26 | 05:41 | 06:24 | 07:12 | 191 |
| 2011 Aug 3 | 01:00 | 01:48 | 02:24 | 181 |
| 2011 Dec 31 | 11:06 | 11:36 | 12:00 | 195 |
| 2012 Mar 30 | 22:43 | 23:24 | 23:48 | 196 |

[a] The position angle is the counterclockwise angle from north to the jet.

Next, we illustrate the characteristics of our jets by presenting for each of two jet events, the EUV jet's eruption observed by the AIA 304 Å and 193 Å movies and the white-light jet's progression in the outer corona observed by LASCO/C2.

### *2.2. First Example Jet*

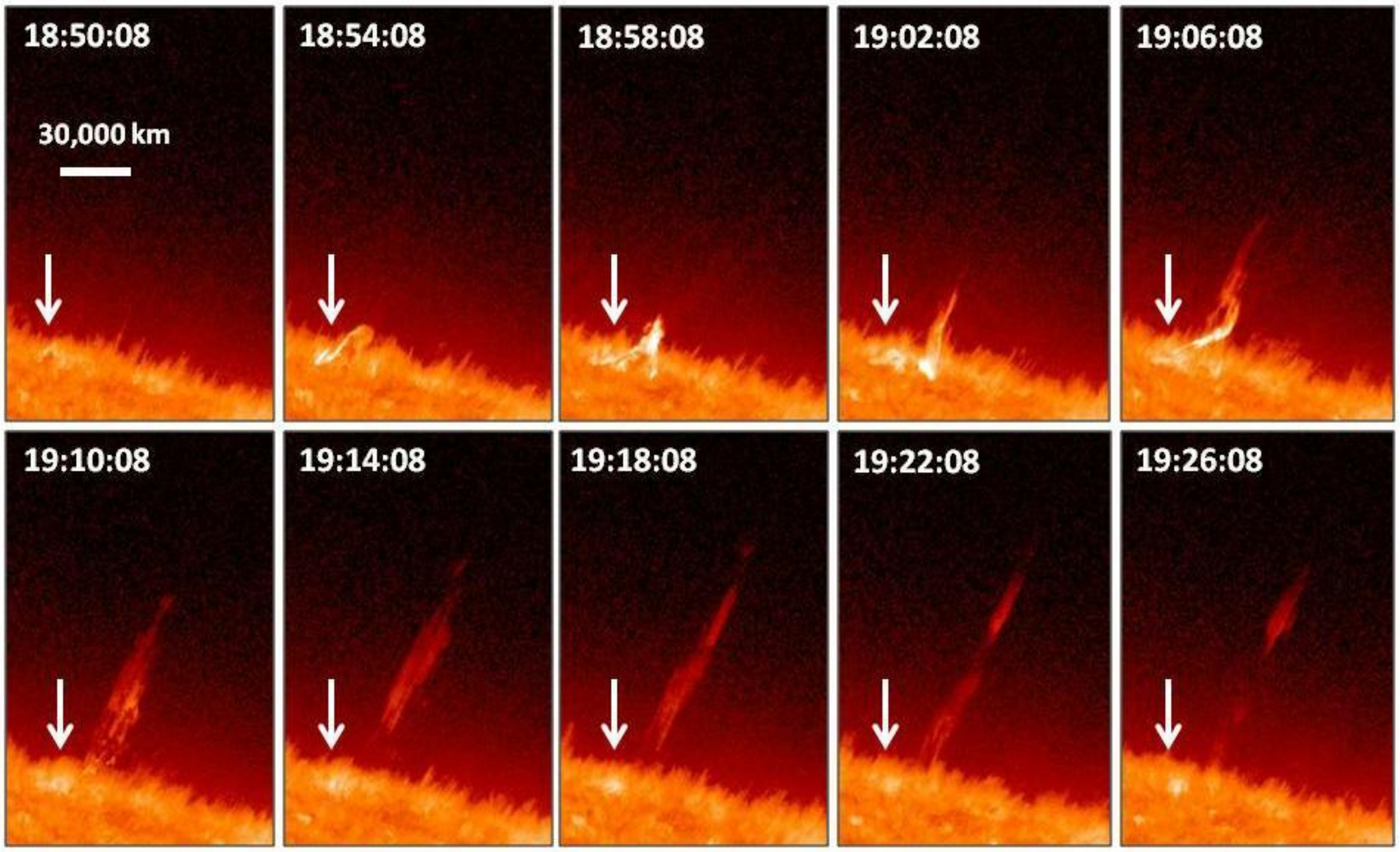

**Figure 1.** The eruption of our first example jet, observed here at 4-minute cadence in images from the *SDO*/AIA He II 304 Å movie. In the upper left corner of each frame is the universal time on 2010 August 11. In these and all other images in this paper solar north is up and west is to the right. (Also see the animation of this figure.)

The two example jets are the fastest of the 14 in our sample. The first of these two occurred in the northern polar coronal hole on 2010 August 11, starting at 18:50 UT. In Figure 1 the cool component of the EUV jet in the jet's eruption and in much of its decay phase is shown at 4-minute cadence by 10 snapshots from the 304 Å movie. Figure 2 shows the jet at the same 10 times in the 193 Å movie. Figures 1 and 2 both show that this jet was produced by blowout eruption of closed field in the base of the jet. In both Figure 1 and Figure 2, the arrow in the first frame points to a compact feature that both movies show is starting to erupt at this time (18:50 UT). The plasma in this feature is evidently at cool-transition-region temperature because it is bright (seen in emission) in the He II 304 Å image and dark (seen in absorption) in the 193 Å image. The 193 Å movie shows that, before it erupts, this feature is low in the core of a lobe of closed field that comprises the eastern half of the base of a coronal plume. This plume is visible in the first frame of Figure 2, and is visible in the 193 Å movie for more than an hour before the start of the jet eruption. The erupting-loop form of this feature in the next two frames of Figures 1 and 2, suggests that this feature is a small filament of cool plasma that is carried in the erupting core of the closed field as in the larger erupting filaments in the blowout eruptions that make CMEs (e.g., see Moore et al 2010). In the 193 Å movie, it appears that the erupting lobe probably drives interchange reconnection high on the outside of its western leg where it impacts ambient open field, starting at about the time of the first frame of Figure 2. The 193 Å jet spire (presumably resulting from the interchange reconnection) becomes faintly visible in the movie by the time of the second frame of Figure 2 (4 minutes later) and is clearly visible by the time of the fourth frame (another 8 minutes later).

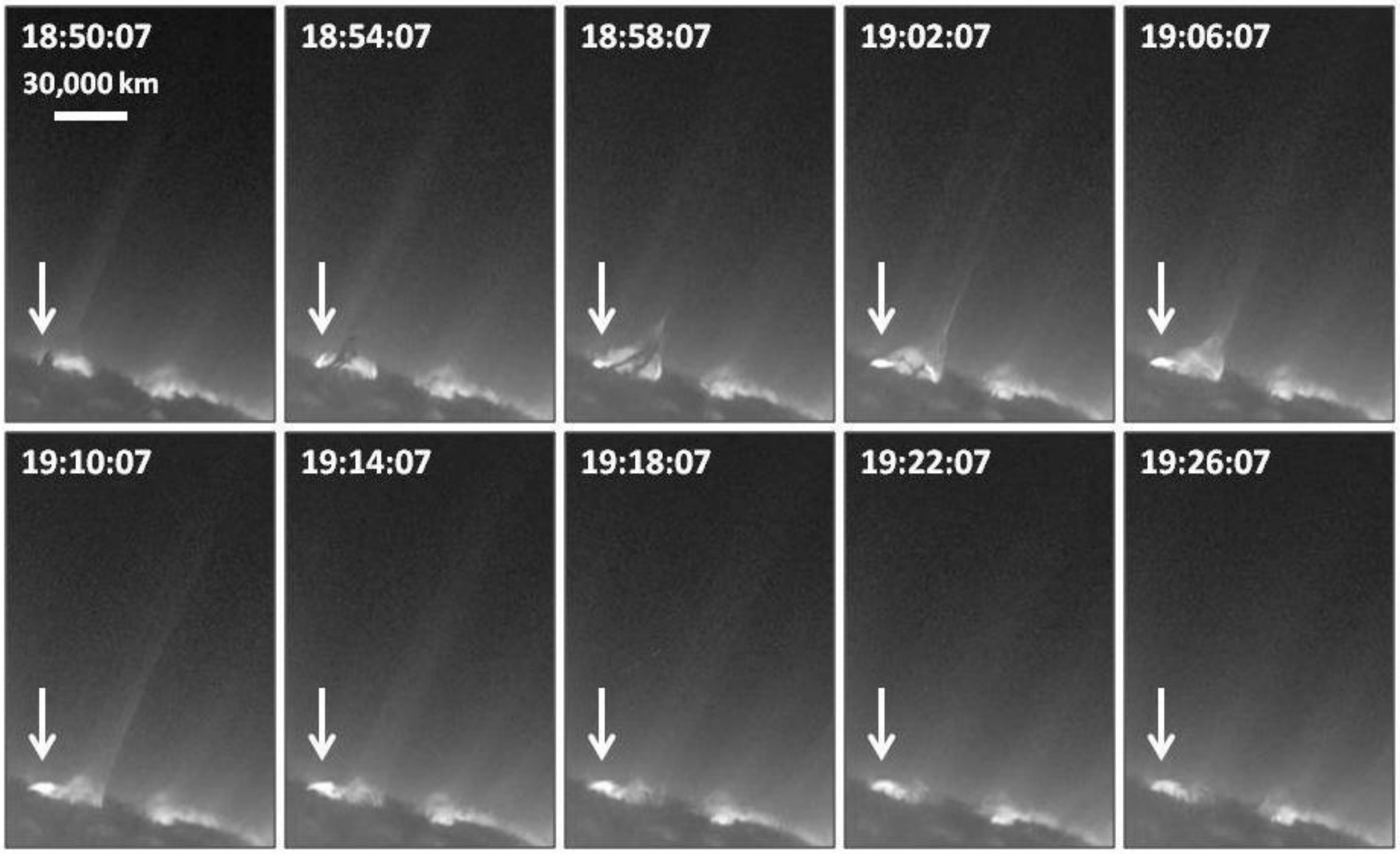

**Figure 2.** The eruption of our first example jet in images from the *SDO*/AIA Fe XII 193 Å movie. The field of view and times of these images match those of Figure 1. (Also see the animation of this figure.)

From both movies, it appears that some of the field threading the filament undergoes interchange reconnection that ejects He II 304 Å filament plasma up along the reconnected open field, and that some of the filament-carrying erupting core field dumps down into the west side of the base of the jet. In the movies, this downfall ends at about the time of the fourth frames of Figures 1 and 2, while another part of the erupting core field, carrying plasma that brightens in both 304 Å and 193 Å emission, continues erupting upward and appears to undergo interchange reconnection that ends at about the time of the sixth frames of Figures 1 and 2. In the Fe XII 193 Å movie, the jet attains its maximum extent at about the time of the sixth frames of Figures 1 and 2, and then shrinks in height and fades to invisibility in the rest of the time covered in Figures 1 and 2. The cool component seen in the He II 304 Å movie continues to reach higher as it fades until about the eighth frames of Figures 1 and 2, and then falls back as it continues to fade.

In this jet, as in all of our jets, the He II 304 Å movie shows the spiraling motion (axial rotation) more clearly than the Fe XII 193 Å movie. Correspondingly, the spiral structure of this jet is fairly obvious in the fifth frame of Figure 1, but is much less noticeable if at all in the fifth frame or any other frame of Figure 2.

In Figure 2, the brightest emission feature in this eruption is a compact arch that forms at the place from which the filament erupts. In both Figure 2 and Figure 1, the arrow in each frame points to that place. Apparently due to obscuration by the foreground He II 304 Å spicule forest, this compact bright feature is not as prominent in the 304 Å images (Figure 1) as in the 193 Å images (Figure 2). The compact bright arch is presumably built from above by internal tether-cutting reconnection in the wake of the filament eruption, as for the flare arcade in the wake of CME-producing filament eruptions (e.g, see Moore et al 2001 and Adams et al 2014). In Figure 2, in frames 4-6, as the jets spire and the compact bright arch continue to grow, a complex arched structure brightens in the jet's base below the spire. This structure spans the base west of the compact bright arch, is not as bright as the compact bright arch, and is a few times larger than the compact arch. The timing, form, position, and size of this larger arched structure are all appropriate for it to have been built from above by interchange reconnection driven by the blowout eruption of the minifilament, as in the new variety of blowout jet recently discovered by Adams et al (2014). From this, together with the occurrence of the compact bright arch at the source of the minifilament eruption, we take the present example jet to be a blowout jet of this kind, driven in the same way by the minifilament eruption as in the jet reported by Adams et al (2014).

Figure 3 shows the jet in the outer corona in a sub field of view of the LASCO/C2 running-difference movie from the CME Catalog. The five frames in Figure 3 are from the five consecutive C2 running-difference images in which the jet was present and bright enough to be discerned. The radius of the outer edge of the occulting-disk mask in these images is 2.2 $R_{Sun}$. The first frame in Figure 3 is from the first C2 image taken after the front of the jet had entered the C2 field of view beyond the mask. In the last frame of Figure 3, the jet has almost faded to invisibility. The black cross in each top frame marks the radial distance given in the CME Catalog for the front of the jet. As is listed in Table 2, according to the CME Catalog, the jet front moved from 3.67 $R_{Sun}$ in the first frame of Figure 3 to 5.97 $R_{Sun}$ in the last frame. The five crosses together in the top row of Figure 3 show that the front of the jet traversed the C2 field of view at a roughly constant speed. From the slope of the least-squares linear fit to these five points, the jet-front speed given in the CME Catalog is 550 km $s^{-1}$. Along most of the jet, with decreasing distance from the front, the jet decreases in brightness until it merges into the background brightness at the front. From this characteristic of the jets in C2 running-difference images such as those in Figure 3, we judge the jet-front radial distances given in the CME Catalog to be uncertain by of order ± 0.1 $R_{Sun}$. (We estimated this uncertainty by visually inspecting the fronts of our jets in the C2 running-difference images. The corresponding uncertainty in the front speeds given for our jets in the CME Catalog is of order ± 10% (see Section 3).)

The image sequence in Figure 3 shows subtle swaying and bending of the jet. The jet leans clockwise from radial in the first frame. Three frames (36 minutes) later, the outer half of the jet still leans clockwise of radial but the inner half leans counterclockwise of radial, so that the jet has an overall camber that is concave to the right-hand side of the jet. In the second frame, the jet subtly displays form that is suggestive of a spiral: with distance beyond the mask, it is first concave to the right and then becomes slightly concave to the left. In Section 6, we take the jet's sway and warp observed in these images to be produced by the magnetic-untwisting wave that is generated along with the jet outflow and that perhaps propels plasma from lower in the corona into the C2 outer corona.

### *2.3. Second Example Jet*

The second example jet happened in the southern polar coronal hole on 2011 April 9, starting at 05:36 UT. Of the 18 polar jets that we found in our search of the LASCO/C2 movies of the outer corona, this jet is the only one that had any coverage in a *Hinode*/XRT movie of a polar coronal hole. It is one of

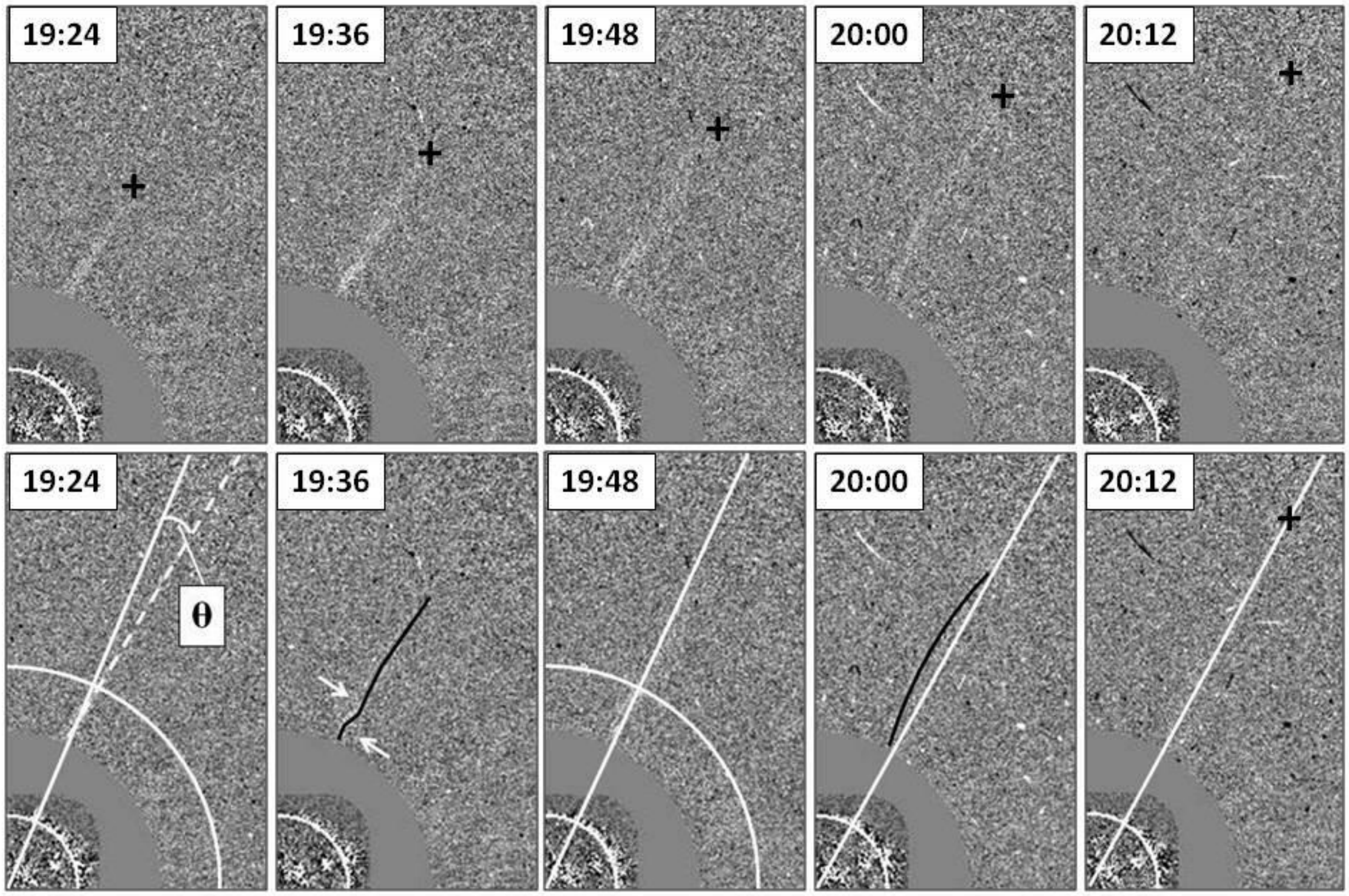

**Figure 3.** The progression and form of our first example jet in the outer corona in *SOHO*/LASCO/C2 running-difference images. The universal time is displayed in the corner of each frame. In each frame, an AIA 193 Å running-difference image is concentric with the occulting-disk mask in the C2 image and is outlined with a white quarter circle showing the edge of the solar disk. Each frame in the top row is repeated in the frame below it with curves and symbols overlaid to point out aspects of the jet that are discernible in the top frame. The first bottom frame delimits the jet's bottom segment (the segment from the 2.2 $R_{Sun}$ outer edge of the gray occulting-disk mask to the 3.0 $R_{Sun}$ quarter circle) and shows that at this time the segment leans clockwise of radial by many degrees. The angle of a jet's bottom segment to the radial direction is the sway angle $\theta$ defined and discussed in Section 5. In the second bottom frame, the black curve (visually drawn to trace the jet's centerline) shows that with distance beyond the occulting-disk mask the jet first appears concave to the west (lower arrow) and then appears slightly concave to the east (upper arrow) before perhaps becoming slightly concave to the west again. In the third bottom frame, the 3.0 $R_{Sun}$ quarter circle delimits the jet's bottom segment, and the radial line through the center of the segment shows that the direction of the segment is now nearly radial. In the fourth bottom frame, the black arc traces the jet's centerline, showing that the entire jet beyond 2.2 $R_{Sun}$ is now concave to the west; the radial line is drawn through the west edge of the jet at 2.2 $R_{Sun}$. Because the jet and its front are hardly visible if at all in the final C2 running-difference image of this jet (fifth top frame), in the fifth bottom frame, the white radial line is repeated from the fourth bottom frame to show that in the fifth top frame, as a reasonable guess, we placed the jet front (black plus) on this radial line at the front's radial distance given for this time in the CME Catalog.

the 32 blowout jets of the 54 polar X-ray jets that Moore et al (2013) studied for characteristics displayed in AIA He II 304 Å movies. That only one of these 54 X-ray jets became in the C2 outer corona a white-light jet that was noticeable enough to be included in the CME Catalog supports our inference in

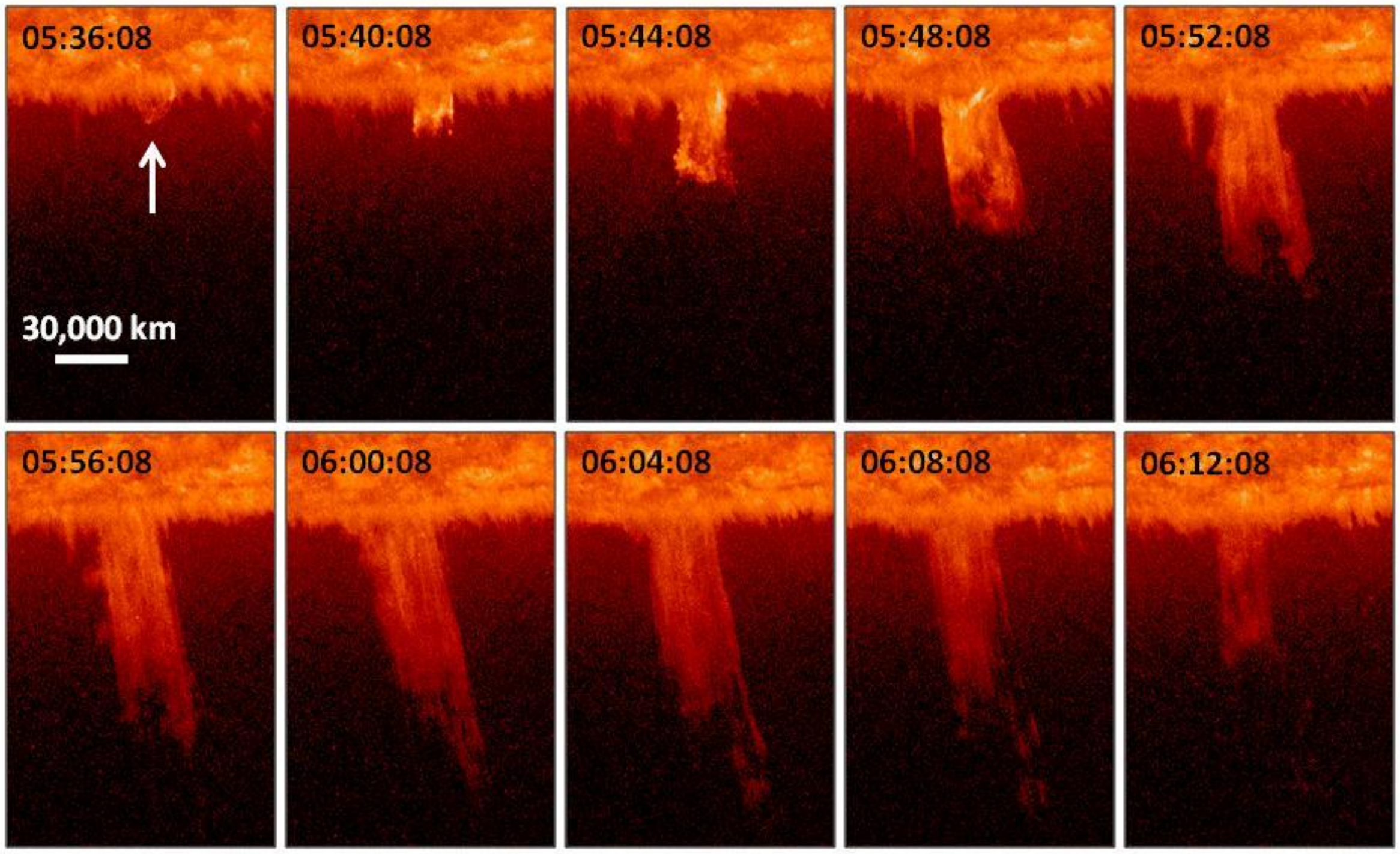

**Figure 4.** The eruption of our second example jet in images from the He II 304 Å movie. The format and cadence are the same as in Figure 1. This jet occurred on 2011 April 9, and erupted from behind the limb in the south polar coronal hole. (Also see the animation of this figure.)

Section 1 that C2 white-light jets such as ours and those studied by Wang et al (1998) come from only certain big polar X-ray/EUV jets.

Figure 4 shows at 4-minute cadence the jet's eruption and evolution via 10 images from the 304 Å movie, and Figure 5 shows the jet's images from the 193 Å movie at the same 10 times. This jet was seated far enough behind the limb that its pre-eruption closed-field base is mostly hidden from view in these movies. The arrow in Figures 4 and 5 points to a closed magnetic loop that is erupting from the base and is beginning to rise above the limb. Only some part of the loop is faintly seen in absorption (dark) in the 193 Å image. Even though the jet is much fainter in the 193 Å movie than in the 304 Å movie, it can be seen in the 193 Å movie that, by the time of the first frame in Figures 4 and 5, the eruption has produced a faint jet spire that extends much higher than the erupting loop, an indication that interchange reconnection has started. Over the next 16 minutes (frames 2-5 in Figures 4 and 5), the blowing-out closed-loop field apparently opens, presumably by interchange reconnection with the ambient open field, and the jet's resulting open-field spray grows wider and taller in the 304 Å movie, while in the 193 Å movie the jet becomes brighter and becomes increasing taller than the cooler component seen in the 304 Å movie. In the 193 Å movie, the jet attains its greatest extent at about the time of frame 6 in Figures 4 and 5, and then fades to invisibility by the time of the last frame, 36 minutes after onset. In the 304 Å movie, the jet spray increases in height until about frame 8 of Figures 4 and 5, and thereafter falls back and fades out.

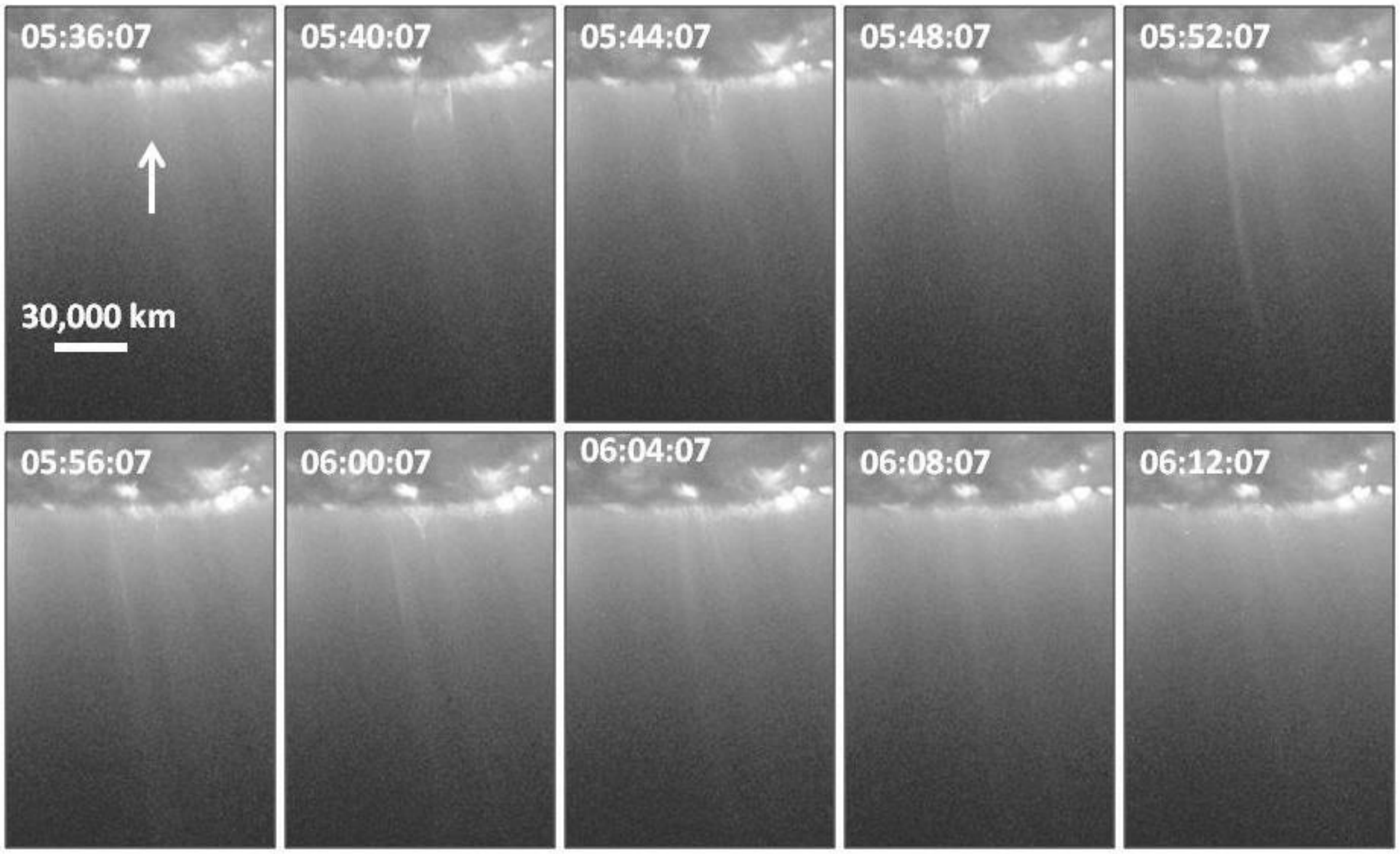

**Figure 5.** The eruption of our second example jet in images from the Fe XII 193 Å movie. This figure corresponds to Figure 4 in the same way as Figure 2 corresponds to Figure 1. (Also see the animation of this figure.)

The axial rotation of the jet is not apparent in the 4-minute-cadence sequence of images in either Figure 4 or Figure 5, but is quite obvious in the 12-second-cadence 304 Å and 193 Å movies. The axial rotation is clearly seen in the 193 Å movie, and the 304 Å movie shows it better yet. The spinning motion begins as the jet erupts and continues throughout the jet's growth, from about the time of frame 1 to about the time of frame 7 in Figures 4 and 5. During this time, the reconnected open field traced by He II 304 Å plasma appears to untwist.

In the same way as Figure 3 for the first example jet, Figure 6 shows the progression and form of the second example jet in the five consecutive C2 running-difference images in which the jet was seen. Again the black cross in each top frame marks the jet front's radial distance given in the CME Catalog. The five crosses together again show that the jet's front moved outward across the C2 field of view at nearly constant speed. The front speed given in the CME Catalog from the least-squares linear fit to these points is 590 km $s^{-1}$.

Figure 6 displays swaying and bending of this jet similar to that displayed by Figure 3 for the first example jet. In frame 1 of Figure 6, the jet leans counterclockwise of radial. In frame 2, the outer half of the jet is still counterclockwise of radial but the lower half is clockwise of radial, so that the jet has overall camber that is concave to the west. In frame 3, the jet has lost its camber and is nearly radial. In frame 4, the lower half of the jet again leans counterclockwise of radial, and the jet has slight overall camber that is concave to the east. In frame 5, the jet is most visible below about 3 $R_{Sun}$ and is seen there to still lean counterclockwise of radial. Again, in Section 6, we take this oscillatory swaying and bending of the jet to be the signature of the jet's co-generated magnetic-untwisting wave: we take the amplitude of the swaying and bending to be the amplitude of the wave as it propagates out through the C2 outer corona.

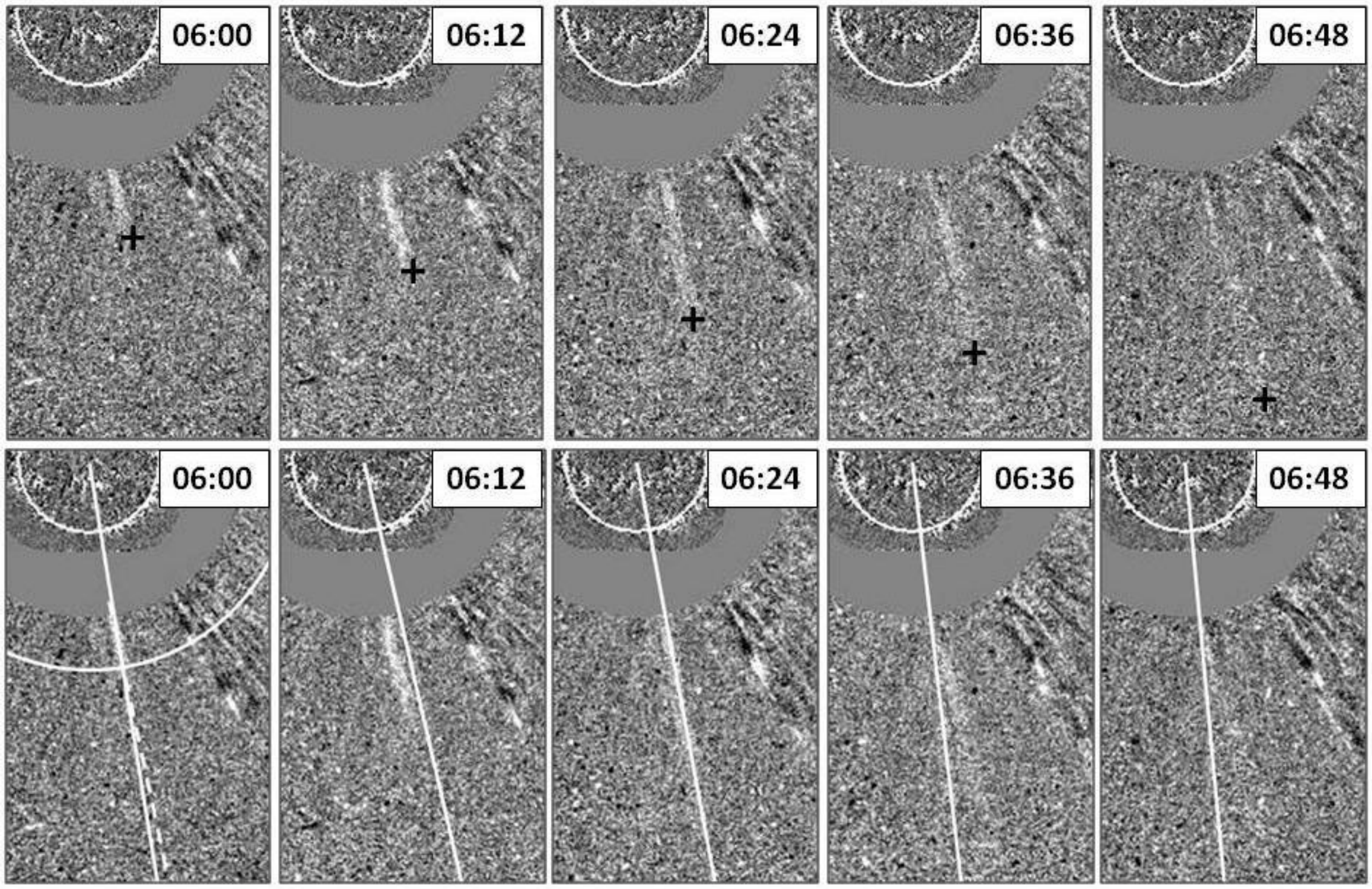

**Figure 6.** The progression and form of our second example jet in the outer corona in LASCO/C2 running-difference images. The format and cadence are the same as in Figure 3. The first bottom frame, in the same way as in the first bottom frame of Figure 3, shows that the jet's bottom segment leans counterclockwise of radial by a few degrees at this time. In the second bottom frame, the radial line is drawn through the west edge of the jet's front to show that the jet's bottom segment now leans clockwise of radial and the jet now has overall camber that is concave to the west. In the third bottom frame, the radial line is drawn through the middle of the jet and shows that the jet now has no discernible camber and is nearly radial. In the fourth bottom frame, the radial line is drawn through the east edge of the jet at 2.2 $R_{Sun}$ to show that the jet's bottom segment now again leans counterclockwise of radial and the jet now has overall camber that is concave to the east. In the fifth bottom frame, the radial line is again drawn through the east edge of the jet at 2.2 $R_{Sun}$ and shows that the jet's bottom segment still leans counterclockwise of radial.

## 3. JET SPEEDS

For each of our 14 jets, as for our two example jets in Figures 3 and 6, the CME Catalog gives the speed of the jet's front given by the slope of the least-squares linear fit to the front's time-distance plot obtained from the jet's sequence of C2 running-difference images. This jet-front speed is listed in the last column of Table 2.

| Table 2<br>Jet-Front Speeds Before and After Jet's First C2 Image | | | | | |
|---|---|---|---|---|---|
| Date | Transit Time (min) | Front's Heliocentric Distance[a] ($R_{Sun}$) In First C2 Image | In Last C2 Image | Jet-Front Transit Speed[b] (km $s^{-1}$) | Jet-Front Speed[a] in C2 Movie (km $s^{-1}$) |
| 2010 Aug 11 | 42 | 2.55 | 5.29 | 430 | 650 |
| 2010 Aug 11 | 34 | 3.67 | 5.97 | 910 | 550 |
| 2010 Aug 19 | 41 | 2.56 | 3.97 | 440 | 340 |
| 2010 Sep 29 | 37 | 2.69 | 4.16 | 530 | 320 |
| 2010 Nov 8 | 51 | 3.01 | 5.04 | 460 | 550 |
| 2010 Nov 14 | 92 | 2.86 | 3.81 | 230 | 300 |
| 2010 Dec 4 | 42 | 2.78 | 3.65 | 490 | 230 |
| 2011 Jan14 | 49 | 2.51 | 3.36 | 360 | 280 |
| 2011 Apr 9 | 25 | 2.94 | 5.36 | 900 | 590 |
| 2011 May 3 | 38 | 3.03 | 5.69 | 620 | 510 |
| 2011Jun 26 | 43 | 2.93 | 4.86 | 520 | 470 |
| 2011 Aug 3 | 48 | 2.93 | 4.36 | 470 | 470 |
| 2011 Dec 31 | 30 | 2.72 | 3.36 | 670 | 310 |
| 2012 Mar 30 | 41 | 3.01 | 4.18 | 570 | 570 |
| Average Values | 44 | 2.87 | 4.50 | 540 | 440 |

[a] From the *SOHO* LASCO CME Catalog (Gopalswamy, et al 2009).

[b] The Jet-front transit speed is the front's mean speed in transiting from the Sun's surface (limb) to where it is in the jet's first C2 image.

From the He II 304 Å movie of the EUV source jet together with the white-light jet's first C2 running-difference image, we can estimate the mean speed of the jet front before the first C2 running-difference image, the front's mean speed in transiting from the Sun's surface to where it is in the jet's first C2 running-difference image. (In all 14 jets but our first example jet, during most of that transit time the front is in the inner corona below the 2.2 $R_{Sun}$ edge of the occulting-disk mask in the C2 images.) This mean speed is approximately the jet-front transit speed that we mentioned in the Introduction, the jet-front speed measured by Wang et al (1998) for their C2 polar jets. The jet-front transit speed that we measured is practically the same as what Wang et al (1998) measured. Ours is the front's distance from the edge of the Sun's disk (the front's distance from disk center minus 1 $R_{Sun}$) in the jet's first C2 running-difference image divided by the front's transit time from the limb to its distance in the jet's first C2 running-difference image (the difference between the jet's first two times in Table 1). The jet-front transit time for each jet is in the second column of Table 2, the front's distance from disk center in the jet's first C2 running-difference image is in the third column, and the jet-front transit speed is in the fifth column.

Comparison of the jets in terms of their C2 front speeds (last column of Table 2) and their persistence in the C2 running-difference movie (the time in the fourth column of Table 1 minus the time in the third column) shows that the faster jets usually last longer in these movies than the slower jets. This is consistent with the tendency seen in Table 2 in the C2 front speeds (fifth column) and final front distances (fourth column): the faster-front jets usually extend farther across the C2 field of view than the slower-front jets before fading out in the C2 running-difference movie. What causes jets having slower fronts to fade out more quickly than jets having faster fronts remains to be found and is beyond the scope of this paper.

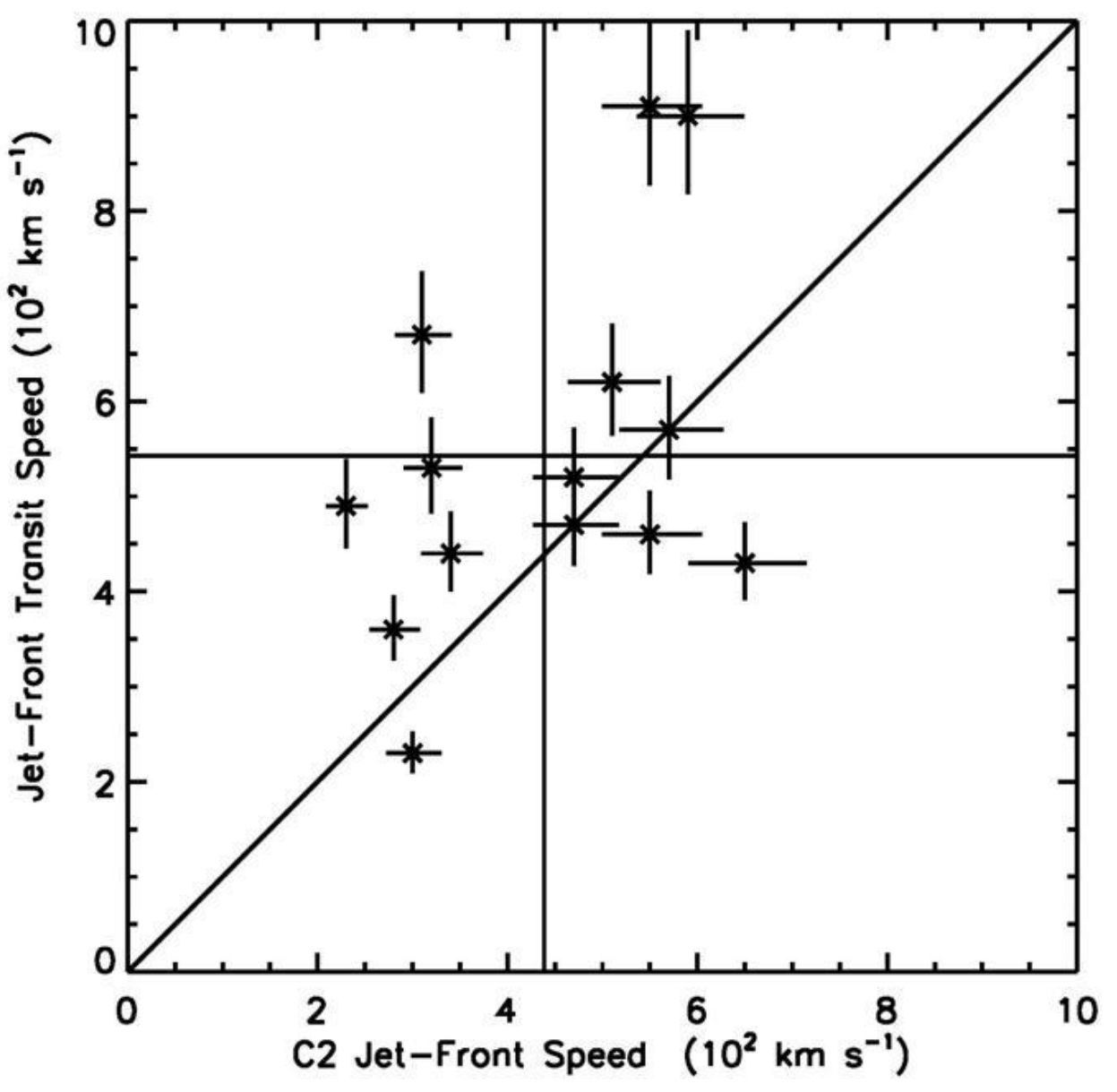

**Figure 7.** Comparison of the jet-front transit speeds of our 14 jets with their C2 jet-front speeds. These speeds are displayed with 10% error bars. The two speeds are equal on the diagonal line. The horizontal line marks the average jet-front transit speed. The vertical line marks the average C2 jet-front speed.

For our 14 jets, the jet-front transit speed in Table 2 ranges from 230 km $s^{-1}$ to 910 km $s^{-1}$, and averages 540 km $s^{-1}$. From the uncertainty in where a jet's front is in a C2 running-difference image and the uncertainty in the time at which a jet starts shooting up from the Sun's surface, we estimate that the measurement uncertainty in either the jet-front transit speeds or the C2 jet-front speeds listed in Table 2 is roughly ± 10%. [We estimate this as follows. The jet-front transit speed is the front's distance beyond 1 $R_{sun}$ in the jet's first C2 running-difference image (Table 2, column 3) divided by the transit time (Table 2, column 2). For our 14 jets, that distance ranges from 1.5 $R_{sun}$ to 2.0 $R_{sun}$ and the uncertainty in each distance is about ± 0.1 $R_{sun}$. So, the uncertainty in the distance ranges from about ± 5% to about ± 7%. The transit time ranges from 25 min to 92 min, and the uncertainty in each transit time is the uncertainty in the start time of the EUV jet, about ± 1 min. So, the uncertainty in the transit time ranges from about ± 1% to about ± 4%. Thus, the uncertainty in the jet-front transit speed is within the range between about ± 5% and about ± 8%. The C2 jet-front speed is approximately the travel distance (the difference of the front's radial distance between the jet's first and last C2 running-difference images (Table 2, columns 3 and 4)) divided by the travel time (the time between these two images). There is negligible uncertainty in that travel time. For our 14 jets, the travel distance ranges from 0.6 $R_{sun}$ to 2.7 $R_{sun}$, and the uncertainty in each travel distance is about ± 0.14 $R_{sun}$. Thus, the uncertainty in the C2 jet-front speed is within the range from about ± 5% to about ± 20%.] We assume that the measurement uncertainty in the jet-front transit speeds measured by Wang et al (1998) is about the same (roughly ± 10%). The slowest, fastest, and mean jet-front transit speeds for the 27 jets measured by Wang et al (1998) are 410 km $s^{-1}$, 1090 km $s^{-1}$, and 630 km $s^{-1}$, marginally significantly faster than for our 14 jets. We think that this difference is probably a jet-selection effect. The cadence of the LASCO/C2 movies was 2-5 times slower during the time that Wang et al (1998) searched than during the time that we searched (20-60 minutes during their search instead of the 12-minute cadence during our search). This probably biased the Wang et al (1998) sample toward longer-lasting white-light polar jets, which tend to be the faster ones.

| Table 3<br>Jet Axial Rotation and Diameter at 1.03 $R_{Sun}$ | | | | |
|---|---|---|---|---|
| Date | Rotation Time Interval (UT) | Number of 360° Turns | Rotation Speed (km $s^{-1}$) | Diameter ($10^3$ km) |
| 2010 Aug 11 | 9:54 – 10:03 | 3/4 ± 1/8 | 60 ± 15 | 22 ± 2 |
| 2010 Aug 11 | 18:57 – 19:13 | 7/4 ± 1/4 | 100 ± 20 | 14 ± 2 |
| 2010 Aug 19 | 20:48 – 20:59 | 5/4 ± 1/8 | 80 ± 20 | 20 ± 2 |
| 2010 Sep 29 | 21:36 – 21:40 | 3/2 ± 1/8 | 200 ± 40 | 10 ± 1 |
| 2010 Nov 8 | 22:10 – 22:27 | 2 ± 1/4 | 100 ± 20 | 18 ± 2 |
| 2010 Nov 14 | 14:04 – 14:18 | 7/4 ± 1/4 | 90 ± 20 | 22 ± 2 |
| 2010 Dec 4 | 21:44 – 21:54 | 7/4 ± 1/4 | 90 ± 20 | 17 ± 2 |
| 2011 Jan 14 | 16:48 – 17:01 | 7/4 ± 1/4 | 110 ± 20 | 19 ± 2 |
| 2011 Apr 9 | 05:37 – 05:59 | 9/4 ± 1/4 | 140 ± 20 | 22 ± 2 |
| 2011 May 3 | 04:15 – 04:25 | 3/4 ± 1/8 | 80 ± 20 | 21 ± 2 |
| 2011 Jun 26 | 05:47 – 06:02 | 7/4 ± 1/4 | 220 ± 40 | 22 ± 2 |
| 2011 Aug 3 | 01:11 – 01:21 | 1 ± 1/8 | 60 ± 15 | 15 ± 2 |
| 2011 Dec 31 | 11:08 – 11:25 | 2 ± 1/4 | 130 ± 20 | 15 ± 2 |
| 2012 Mar 30 | 22:50 - 23:01 | 5/4 ± 1/8 | 120 ± 20 | 26 ± 3 |
| Average Values | 13 min | 1.5 | 110 | 19 |

For our 14 jets, the transit speed of the front was usually significantly faster than the front's speed in the C2 outer corona: in Table 2, the average jet-front transit speed is 540 km $s^{-1}$ whereas the average jet-front speed from the C2 movies is 440 km $s^{-1}$. The plot of jet-front transit speed versus C2 jet-front speed is shown in Figure 7 with 10% error bars. This plot shows that the two speeds are correlated: jets with faster jet-front transit speeds tend to have faster speeds in the C2 outer corona. Figure 7 also shows that for 9 of the 14 jets the jet-front transit speed was greater than the front's speed in the C2 outer corona. This suggests that the jet-front plasma usually decelerates as it transits the inner corona, resulting in its speed in the C2 outer corona being less than its speed in the inner corona (e.g., Ko et al 2005). Another possibility, one that is suggested by our results in Section 6, is that the magnetic-untwisting wave that is co-generated with the EUV jet imparts an upward speed to the plasma it propagates through that is less than the wave's propagation speed (in this case, the jet-front plasma does not come from the base of the corona but from progressively higher along the path of the wave and the jet). This might give a jet-front transit speed that is faster than the jet-front speed from the C2 movie, e.g., if the Alfven speed in polar coronal holes is somewhat faster in the inner corona below 2.2. $R_{Sun}$ than in the C2 outer corona, which it plausibly can be (Moore et al 1991).

## 4. AXIAL ROTATION OF COOL COMPONENT

Each of our 14 EUV jets displayed axial rotation, spin about the jet's axis, as the jet erupted. In each jet, the axial rotation can be tracked better in the 304 Å movie than in the 193 Å movie (as was mentioned in Section 2, this is due to the better visibility of jet structure in the 304 Å movie than in the 193 Å movie). By stepping through the 304 Å movie, we counted for each jet the total number of 360° turns that the He II 304 Å cool component underwent. This was done in the same way as in Moore et al (2013), by visually tracking features in the periphery of the jet spire across the face of the spire (as in Figures 8-10 of Moore et al (2013) and as in Figures 8 and 9 of the present paper). A prominent feature was tracked across the face from one edge of the spire to the other through the first ½ turn, then another feature was tracked through the next ½ turn, and so forth through the last fraction of a half turn of the spire's rotation. In this way we estimated both the number of turns the spire underwent and the uncertainty in that number. For our 14 jets, the uncertainty in the discernible number of turns ranged from ± 1/8 to ± 1/4 turn. For each jet, the observed number of turns and its uncertainty are in the third column of Table 3. The time interval from the beginning to the end of each jet's rotation is in the second column. For our 14 jets, the duration of the axial rotation ranges from 4 to 22 minutes, and the average duration of rotation is 13 minutes. The number of turns of axial rotation ranges from 3/4 to 9/4 turn, and the average is 1.5 turn.

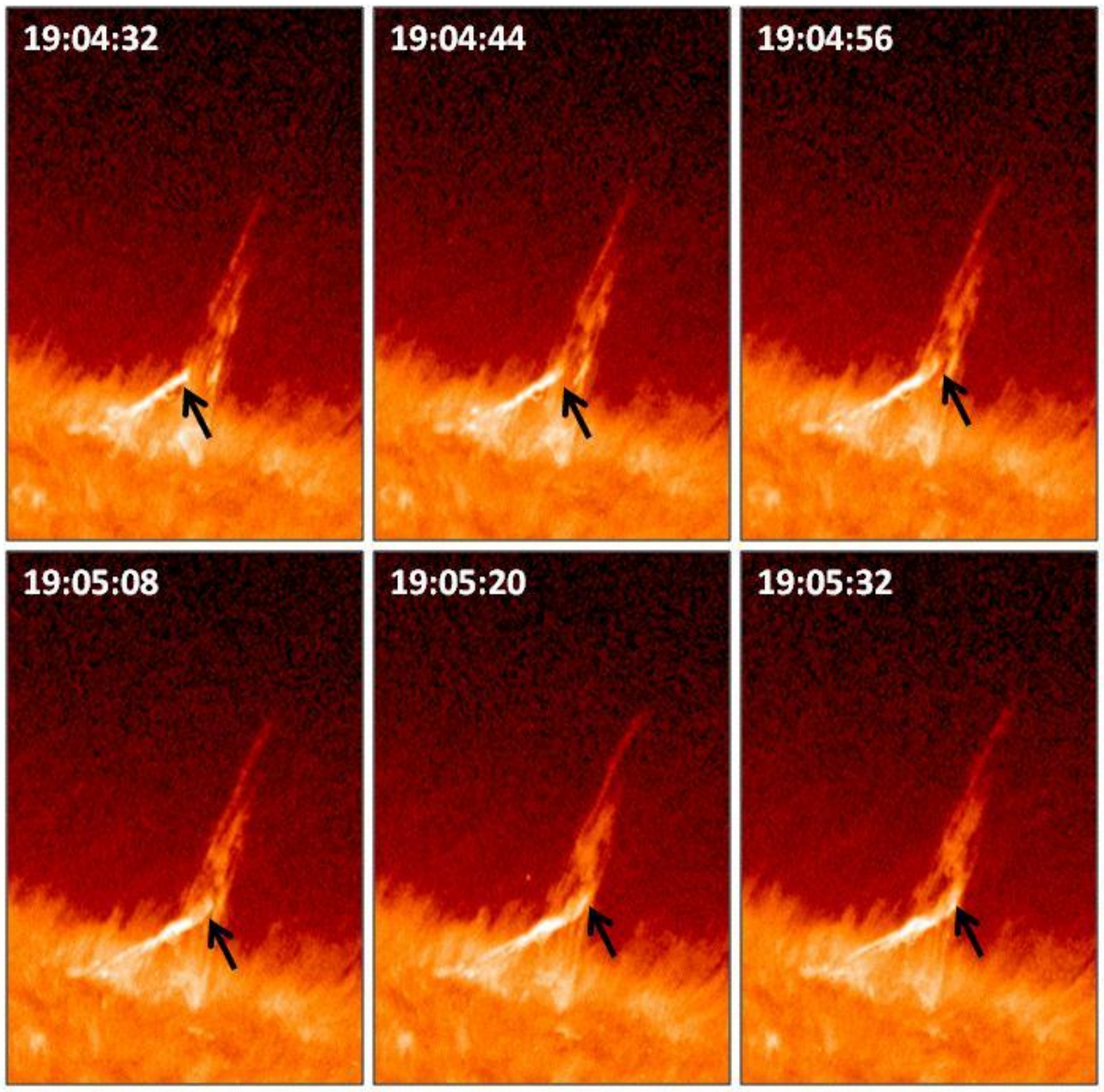

**Figure 8.** Display of the axial rotation of our first example jet in a 1-minute interval during the jet's eruption. The universal time is in the corner of each frame.

Near the middle of the axial-rotation time interval of each jet, we measured the axial-rotation speed of the outside of the cool-component spire by tracking a substructure feature, a strand or clump of plasma, on the face of the spire in 6 consecutive frames of the 304 Å movie. The measured speed is the measured distance the feature moved orthogonal to the axis of the spire divided by the time interval, 60 s. Figure 8 shows the 6 consecutive 304 Å images from which the rotation speed was measured for our first example jet, the jet shown in Figure 1, and Figure 9 shows the 6 images from which the rotation speed was measured for our second example jet, the jet shown in Figure 4. The arrows in Figures 8 and 9 point to the feature that we tracked to measure the rotation speed. For each of our 14 jets, the measured rotation speed and its uncertainty are in the fourth column of Table 3. The rotation speeds range from 60 to 220 km $s^{-1}$ and average 110 km $s^{-1}$. The estimated uncertainty in a jet's rotation speed is entirely from the estimated uncertainty in the distance orthogonal to the jet's axis the tracked feature moved in the 60 s it was tracked, and is typically about ± 20% (Table 3). [The AIA 304 Å images have solar north up and solar west right: the pixel columns run north-south and the pixel rows run east-west. The distance $d_{orth}$ traveled orthogonal to the jet's axis is given by $d_{east\text{-}west}/\cos\alpha$, where $d_{east\text{-}west}$ is the east-west distance traveled and $\alpha$ is the acute angle of the jet axis to solar north-south. For our measurements, the bulk of the uncertainty in $d_{orth}$ is from the uncertainty in $d_{east\text{-}west}$; less than 10% of the uncertainty is from the uncertainty in $\alpha$. For instance, consider our measurement of the rotation speed for our second example jet (the jet of 2011 April 9 in Table 3), from the images shown in Figure 9. We measured $d_{east\text{-}west}$ by counting the number of east-west pixels the tracked feature moved accoss during the 60 s shown in Figure 9. It crossed 19 ± 3 east-west pixels, the uncertainty being due to the changing form and east-west extent of the tracked feature. The pixel size (0.6 arcsec or 435 km on the Sun) and the 60 s time interval give [(138 ± 22)/cos $\alpha$] km $s^{-1}$ for the rotation speed. From Figure 4, we estimate that during the time of Figure 9 the angle $\alpha$ of the jet's axis to north-south was somewhere in the range from about 0° to about 15°. For $\alpha$ = 0°, cos $\alpha$ =1, giving the rotation speed and its uncertainty to be at least about 138 ± 22 km $s^{-1}$. For $\alpha$ = 15°, cos $\alpha$ = 0.966, giving the rotation speed and its uncertainty to be at most about 143 ± 23 km $s^{-1}$. Rounding to the nearest 10 km $s^{-1}$ gives the 140 ± 20 km $s^{-1}$ listed for the rotation speed of this jet in Table 3.]

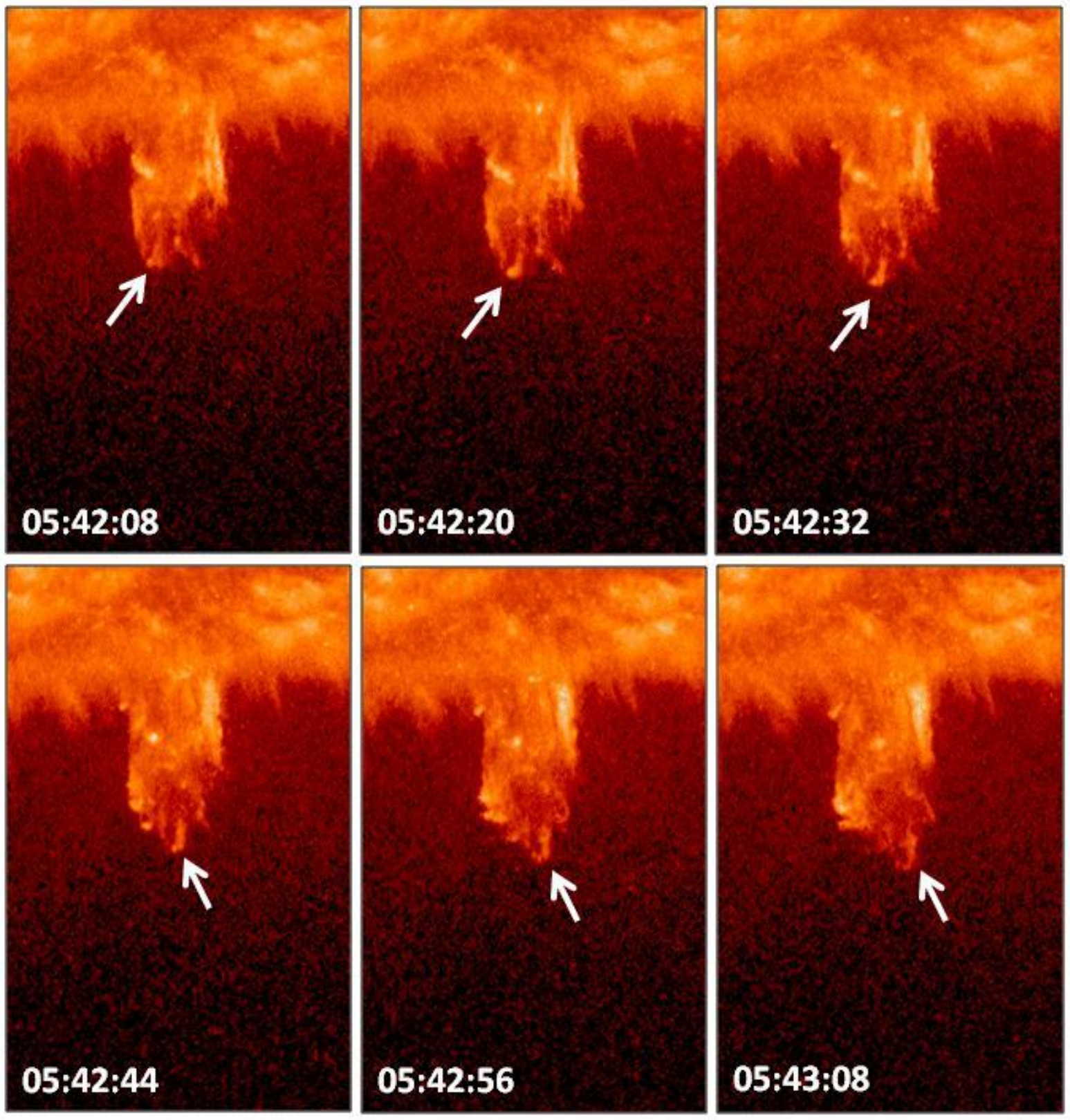

**Figure 9.** Display of the axial rotation of our second example jet in a 1-minute interval during the jet's eruption. The universal time is in the corner of each frame.

For each jet, at the time of the measured speed of axial rotation, we also measured the diameter of the jet's He II 304 Å cool-component spire at a height around 20,000 km (≈ 1.03 $R_{Sun}$ from Sun center). For each jet, the measured diameter and its uncertainty are in the last column of Table 3. The diameters range from 10,000 km to 26,000 km and average 19,000 km. The uncertainty in the measured diameter is from the uncertainty in the locations of the two side edges of the jet spire (due to the raggedness of the edges) in each of the six 304 Å images used to measure the jet's rotation speed, and from changes in location of each edge among the six images (e.g., see the images is Figures 8 and 9). For each jet, by visual inspection of these variations in the spire edges in the six consecutive 304 Å images of the jet, we estimated the uncertainty in the measured diameter to be roughly ± 10 % (see Table 3).

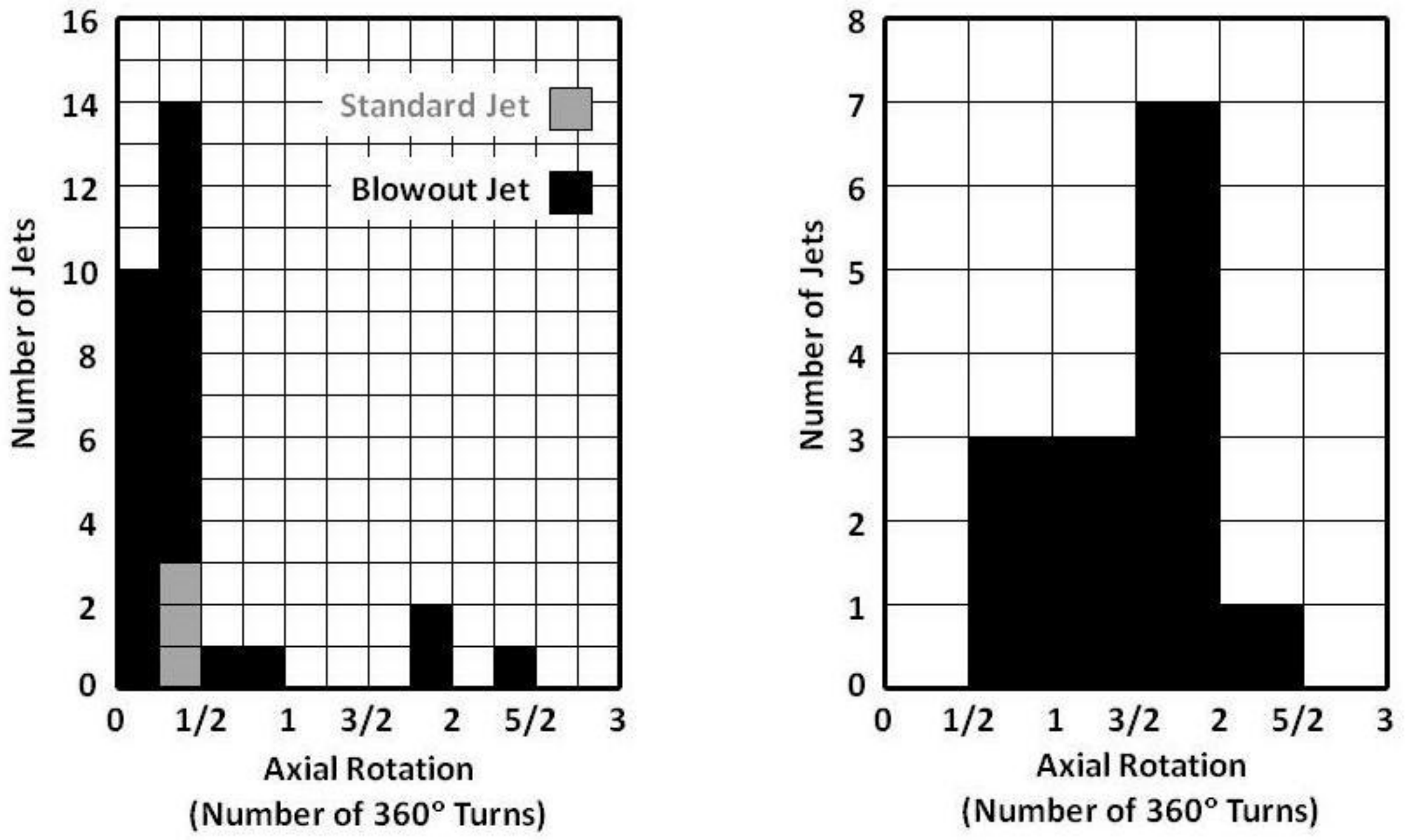

**Figure 10.** Histograms of axial rotation of polar coronal jets. Left: The axial-rotation distribution of 29 random X-ray jets of the size of the 14 EUV jets of the present paper (from Moore et al 2013). Right: The axial-rotation distribution of the 14 EUV jets of the present paper.

Jet-spire transient helical structure and untwisting have been reported in several previous studies of large EUV jets in polar coronal holes (e.g., Wang et al 1998; Patsourakos et al 2008; Raouafi et al 2010; Sterling et al 2010a), but to our knowledge the number of turns of unwinding counted by feature tracking in the way done here has been previously reported only in Moore et al (2013). [Shen et al (2011) estimated the number of turns of untwisting of a large EUV polar blowout jet (by multiplying the measured angular rotation rate by the observed duration of rotation) to be in the range 1.2 to 2.6. By the same method as Shen et al (2011), Chen et al (2012) estimated the number of turns of untwisting in another large EUV blowout jet in a polar coronal hole to be about 3.6. Also by the same method as Shen et al (2011), for another large EUV polar blowout jet, one that was observed to go into the outer corona beyond 1.5 $R_{Sun}$ by the STEREO/COR1 coronagraph, Hong et al (2013) estimated the number of turns of untwisting to be about 0.9.] On the right in Figure 10 is a histogram of the turns of axial rotation that we measured from the He II 304 Å cool component of our 14 EUV jets, all of which were blowout jets. For comparison, on the left in Figure 10 is a histogram of the turns of axial rotation that Moore et al (2013) measured from the He II 304 Å cool component of 29 random polar X-ray jets, of which 26 were blowout jets and 3 were standard jets. These X-ray jets were about the same size as our 14 EUV jets that became white-light jets observed beyond 2.2 $R_{Sun}$ by LASCO/C2. The axial-rotation histogram for the Moore et al (2013) jets (Figure 10, left) shows that a large majority polar coronal jets of this size (24 of 29) have axial rotation of 1/2 turn or less. In contrast, all of our 14 jets had more than 1/2 turn of axial rotation (Table 3 and Figure 10, right). Thus, Figure 10 indicates that polar jets that reach into the LASCO/C2 outer corona usually have exceptionally large axial rotation. This suggests that polar jets having more axial rotation usually extend to greater heights than polar jets having less axial rotation.

**Table 4**
Jet Diameter and Sway Sweep at 2.3 $R_{Sun}$

| Date | Diameter ($10^3$ km) | Sway Sweep (degrees) |
|---|---|---|
| 2010 Aug 11 | 120 ± 10 | 12.0 ± 1.5 |
| 2010 Aug 11 | 150 ± 20 | 13.0 ± 1.5 |
| 2010 Aug 19 | 120 ± 10 | < 1.0 |
| 2010 Sep 29 | 100 ± 10 | 8.5 ± 1.5 |
| 2010 Nov 8 | 120 ± 10 | 2.5 ± 1.5 |
| 2010 Nov 14 | 180 ± 20 | 4.5 ± 1.5 |
| 2010 Dec 4 | 96 ± 10 | 2.5 ± 1.5 |
| 2011 Jan 14 | 110 ± 10 | < 1.0 |
| 2011 Apr 9 | 120 ± 10 | 8.5 ± 1.5 |
| 2011 May 3 | 150 ± 10 | 3.0 ± 1.5 |
| 2011 Jun 26 | 120 ± 20 | 5.0 ± 1.5 |
| 2011 Aug 3 | 120 ± 20 | 5.5 ± 1.5 |
| 2011 Dec 31 | 120 ± 10 | 3.0 ± 1.5 |
| 2012 Mar 30 | 120 ± 10 | 1.0 ± 1.0 |
| Average Values | 125 | 5 |

As was stated in Section 2.1, from the *SOHO* LASCO CME Catalog we found a total of 18 LASCO/C2 white light jets for which the source EUV jet was observed in *SDO*/AIA EUV movies. Of these 18 EUV jets, only two appeared to be standard jets; the rest were evidently blowout jets. This paucity of standard-jet sources of jets observed in the outer corona by LASCO/C2, together with the axial-rotation histograms in Figure 10, suggests the reason that far fewer LASCO/C2 white-light jets come from EUV standard jets than from EUV blowout jets might be that far fewer EUV standard jets than EUV blowout jets have exceptionally large axial rotation.

## 5. SWAY OF JET IN OUTER CORONA

For the great majority of our jets (12 of 14), the sequence of LASCO/C2 running-difference images of each jet showed perceptible swaying/bending of the jet similar to that seen in the image sequences shown in Figures 3 and 6 for the two example jets. We measured the amount that each jet swayed in the C2 outer corona by measuring in each C2 running-difference image of the jet the angle $\theta$ between the radial direction and the bottom segment of the jet, the segment from the edge of the occulting-disk mask at 2.2 $R_{Sun}$ out to 3.0 $R_{Sun}$. We chose to measure the direction of this bottom segment of each jet because, over the life of a jet in the C2 running-difference images, usually the jet was brighter and showed its direction more clearly in this segment than in its extent beyond 3.0 $R_{Sun}$, as in the two example jets in Figures 3 and 6. From a print of each C2 running-difference image of a jet, we found the angle $\theta$ of the jet's bottom segment by drawing with a straightedge two lines through the 2.6 $R_{Sun}$ midpoint of the segment, as illustrated in the bottom first frame of Figure 3. One line, the solid line in the bottom first frame of Figure 3, is the heliocentric radial line through the segment's midpoint. The other line, the dashed line in Figure 3, is the straight line that, as closely as can be visually discerned, is in the direction of (i.e., tangent to) the segment's centerline at the midpoint. The measured angle $\theta$ is the angle between these two lines. By this visual fitting procedure, we estimated that the uncertainty in a segment's direction angle $\theta$ obtained in this way is about ± 1°, taking into consideration that, by using a protractor, we measured $\theta$ with a precision of about ±0.5°.

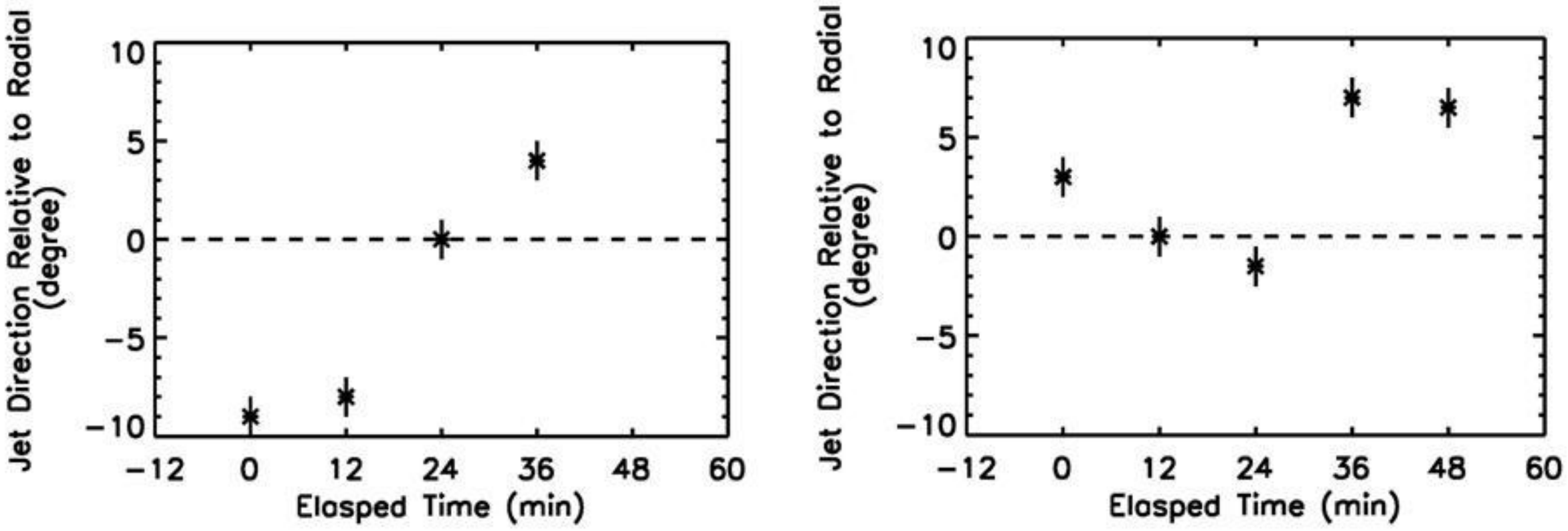

**Figure 11.** Progression of the sway in our two example jets. Left: the example jet in Figure 3. Right: the example jet in Figure 6. Each point shows the angle of the axis of the segment of the jet below 3.0 $R_{Sun}$ in the C2 running-difference image of that time, the angle to the radial direction. This angle is negative, 0, or positive when the direction of this segment is clockwise of radial, radial, or counterclockwise of radial. The error bars are ± 1°.

For the example jet in Figure 3, the jet's segment below 3.0 $R_{Sun}$ in each of the first four of the jet's five C2 running-difference images is bright enough for its direction to be measured, but not in the fifth image. The progression of the measured angle $\theta$ in this jet is plotted on the left in Figure 11. This plot shows that during the 36 minutes spanned by the four measured images, the jet's segment below 3.0 $R_{Sun}$ swayed counterclockwise, from 9° clockwise of radial to 4° counterclockwise of radial. The four plotted points suggest that the swaying was oscillatory, roughly sinusoidal in time with a period of roughly 90 minutes, the four images spanning about 10 minutes less than half a period. That the oscillation is not centered on the radial direction ($\theta$ = 0) but offset a few degrees clockwise of radial is consistent with the direction of the open magnetic field guiding the jet being clockwise of radial due to this field being rooted west of the center of the northern polar coronal hole.

For the example jet in Figure 6, in all five of the jet's C2 running-difference images the jet's segment below 3.0 $R_{Sun}$ is bright enough for its angle $\theta$ to be measured. The progression of the measured $\theta$ is plotted on the right in Figure 11. This plot shows that during the 48 minutes spanned by the five images, the jet's segment below 3.0 $R_{Sun}$ swayed first clockwise and then counterclockwise, indicating that the swaying was oscillatory with a period of roughly 60 minutes, the five images spanning about 10 minutes less than a whole period. In agreement with the open magnetic field guiding this jet being rooted west of the center of the southern polar coronal hole, the oscillation is about a direction that is a few degrees counterclockwise of radial.

For each of our 14 jets, we measured the angle $\theta$ from the jet's C2 running-difference images and obtained a plot of the progression of $\theta$ such as the plots in Figure 11. From each jet's $\theta$ plot, we obtain the jet's observed range of $\theta$, $\Delta\theta$, which we name the jet's "sway sweep." From the plots in Figure 11, the sway sweep of the first example jet is 13.0° ± 1.5° [(4° ± 1°) – (-9° ± 1°)], and the sway sweep of the second example jet is 8.5° ± 1.5° [(7° ± 1°) – (-1.5° ± 1°)]. In addition to obtaining each jet's $\Delta\theta$ by measuring $\theta$ from all of the jet's C2 running-difference images from which $\theta$ could be measured, we also measured the jet's diameter at 2.3 $R_{Sun}$ (0.1 $R_{Sun}$ beyond the edge of the occulting-disk mask) in the jet's first C2 running-difference image. The uncertainty in this measured diameter is about ± 10%.

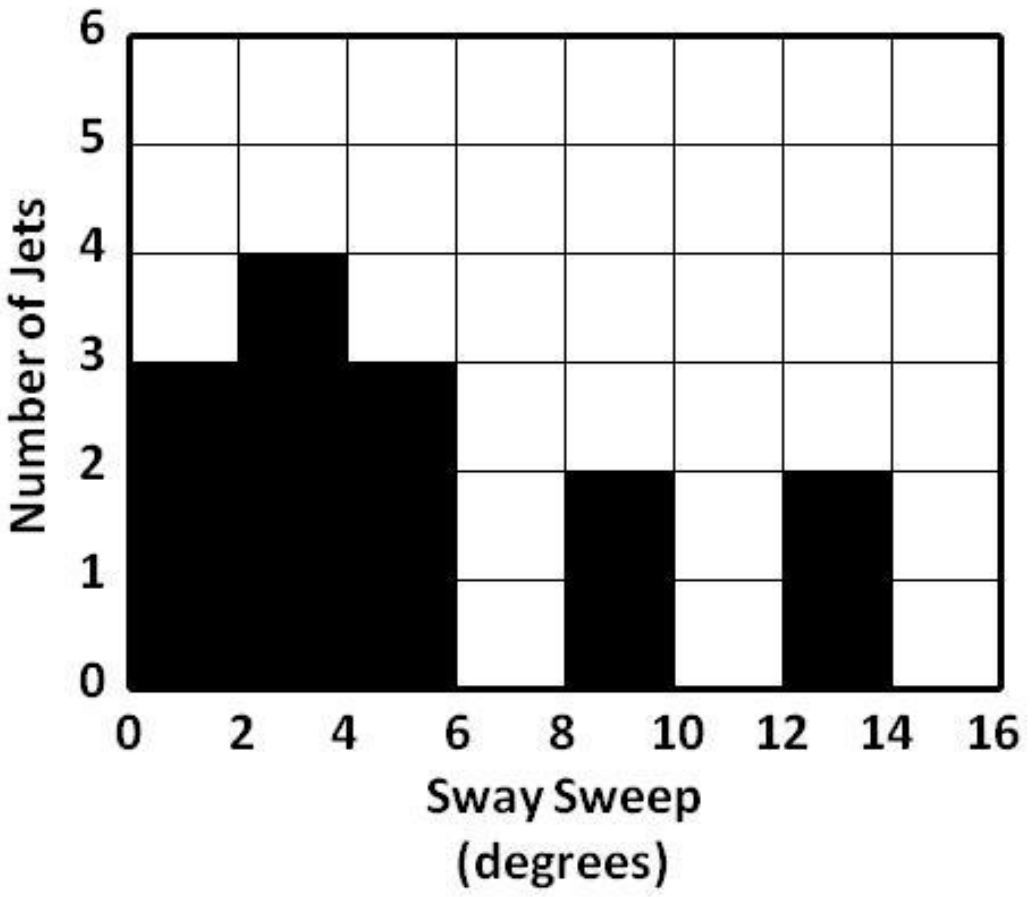

**Figure 12.** Histogram of the sway sweep of our 14 jets.

Table 4 gives the diameter and sway sweep measured for each of our 14 jets. This shows that the jet diameters at 2.3 $R_{Sun}$ ranged from 96,000 km to 180,000 km and averaged 125,000 km. The sway sweep ranged from < 1.0° to 13.0° and averaged 5°. A histogram of the sway sweep of the 14 jets is shown in Figure 12. Table 4 and this histogram show that our two example jets had exceptionally large sway sweep. At 2.3 $R_{sun}$, our first example jet is the wider of the two jets that happened on 2010 August 11. As Table 4 shows, this jet, which had the fastest jet-front transit speed (Table 2), had the largest sway sweep of the 14 jets. Our second example jet (2011 April 9), which had the second fastest jet-front transit speed (Table 2), had the third largest sway sweep of the 14 jets.

That each of our two example jets have both exceptionally fast jet-front transit speed and exceptionally large sway sweep (1) suggests that the magnetic-untwisting wave may be the driver of the jet plasma outflow, and (2) is compatible with our assumption in Section 6 that the sway sweep shows the amplitude of the magnetic-untwisting wave in the C2 outer corona.

## 6. INTERPRETATION

### *6.1 Magnetic Driving*

At 1.03 $R_{Sun}$ (about 20,000 km above the photosphere) in polar coronal holes, the magnetic field is open, its strength B is about 10 G (Ito et al 2010), and it fans out faster than radial with increasing radial distance. The electron number density $n_e$ at 1.03 $R_{Sun}$ in polar coronal holes is about 1 x $10^8$ $cm^{-3}$ (Allen 1973). So, the Alfven speed $V_A$ there is about 2000 km $s^{-1}$. ($V_A = (4\pi\rho)^{-1/2}$ B, where $\rho$ is the plasma mass density; for fully ionized solar coronal plasma, $\rho \approx 2.0 \times 10^{-24}$ $n_e$ gm $cm^{-3}$, and $V_A \approx 2.0 \times 10^{11}$ $n_e^{-1/2}$ B cm $s^{-1}$.) Above 1.03 $R_{Sun}$, out to about 6 $R_{Sun}$ in polar coronal holes, the plasma temperature T (mean temperature of the electrons and protons) is about 1 x $10^6$ K (Withbroe 1988), and the sound speed $V_s$ is about 150 km $s^{-1}$. ($V_s = (\gamma kT/m)^{1/2}$, where $\gamma$ is the ratio of specific heats, k is the Boltzmann constant, and m is the mean particle mass; for fully ionized coronal plasma, $\gamma = 5/3$, m ≈ 1.0 x $10^{-24}$ gm, and $V_s \approx 1.5 \times 10^4$ $T^{1/2}$ cm $s^{-1}$.) For typical rates of super-radial lateral expansion of the open magnetic field in polar coronal holes, the Alfven speed is roughly constant (gradually varies by less than a factor of 2) with distance between 1.03 $R_{Sun}$ and 3 $R_{Sun}$ (Moore et al 1991).

In our 14 jets, the jet front moved faster than the ambient sound speed but slower than the ambient Alfven speed. Among the 14 jets, the slowest, fastest, and average jet-front transit speeds were 230 km $s^{-1}$, 910 km $s^{-1}$, and 540 km $s^{-1}$, and the slowest, fastest, and average C2 jet-front speeds were 230 km $s^{-1}$, 650 km $s^{-1}$, and 440 km $s^{-1}$ (Table 2). If the magnetic field in the jet outflow were inactive in the propulsion of the jet-front plasma, but only passively guided the outflow, then this plasma would have been propelled only by the gradient of its own pressure pushing it out along the guide field. The resulting speed of the jet front could then be no greater than of order the sound speed in the outflow. A jet-front speed of 500 km $s^{-1}$ would require the jet plasma to have been heated to $T \sim 1 \times 10^7$ K. The temperature of the out-flowing plasma in EUV/X-ray polar jets of the size of ours has been measured to be about $2 \times 10^6$ K (Pucci et al 2013), a factor of 5 less than needed. The electron density at heights around 20,000 km in EUV/X-ray polar jets of the size of ours has been measured to be about $6 \times 10^8$ $cm^{-3}$ (Pucci et al 2013), from which, with a field strength of 10 G, the Alfven speed at 1.03 $R_{Sun}$ in a jet is about 800 km $s^{-1}$, which is close to the front speeds of our fastest jets. From these considerations, we propose that the plasma in our jets in the C2 outer corona was somehow propelled to 2.2 $R_{Sun}$ and beyond mainly via the magnetic field in the jet. The obvious prospective magnetic driver suggested by the observed motions of our jets is the magnetic-untwisting wave that is manifested by the observed axial rotation and that is plausibly generated by interchange reconnection of ambient open field with initially-closed twisted field erupting from the base of the jet.

### *6.2. Prospective Driver: Magnetic-Untwisting Wave*

#### *6.2.1. Scenario*

Each of our 14 jets was a blowout jet. Figure 13 is our schematic of the generation of the magnetic-untwisting wave in blowout jets. We have tailored these cartoons to mimic the location and size of the compact bright arch relative to the spire in our first example jet (Figures 1, 2, and 8). The four cartoons show four sequential stages of a blowout jet from beginning to end.

The upper-left cartoon shows the magnetic field setup before or during the start of the eruption, before the start of interchange reconnection of the erupting closed field with the encountered ambient open field. The black horizontal line represents the top of the photosphere at the feet of the magnetic field. The two plus signs label positive-polarity magnetic flux domains, and the minus sign labels the domain of negative-polarity flux of two closed bipolar magnetic fields that are embedded in the positive-polarity open field of a coronal hole. The sheared-core closed-field bipole that blows out encloses the polarity inversion line (viewed end-on) on the left, and the somewhat larger closed bipole that will have more closed field added to it by the interchange reconnection encloses the polarity inversion line on the right. The curled field line depicts the blowout bipole's twisted flux-rope core (viewed end-on) that is poised to erupt or has started to erupt. The short black line segment between the flux rope and the open field marks the current sheet at which interchange reconnection will occur. The short black line segment between the legs of the flux rope marks the current sheet at which internal tether-cutting reconnection will occur as in the larger blowout eruptions that make a flare arcade in tandem with a CME.

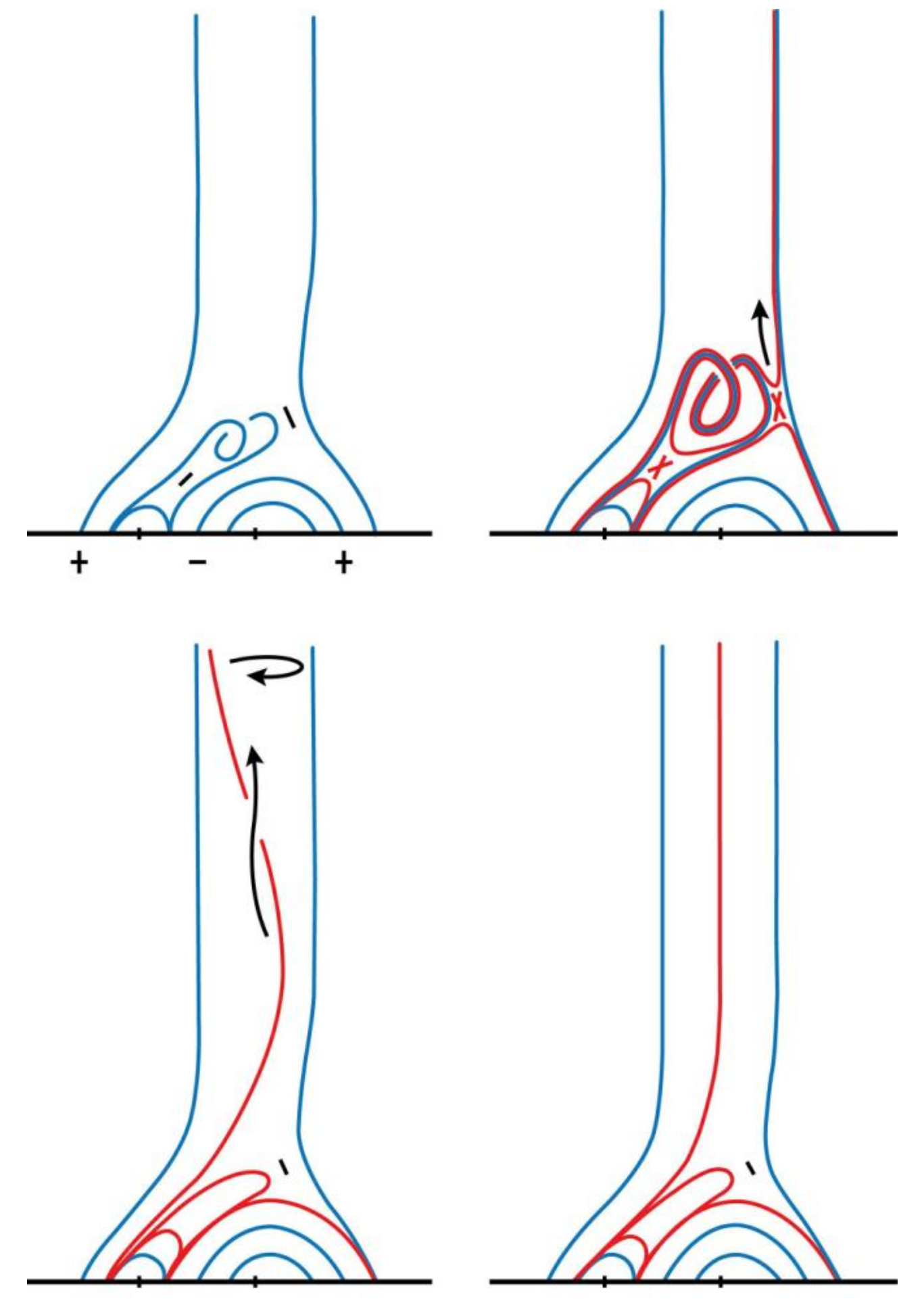

**Figure 13.** Schematic of the generation of the magnetic-untwisting wave in a blowout jet by the blowout and interchange reconnection of magnetic field at the base of the jet. Only a few rudimentary field lines are drawn to convey the essentials of the form and action of the magnetic field.

The upper-right cartoon shows the explosive growth phase of the blowout jet eruption. Each of the two places at which reconnection is occurring is marked by a red X. Field lines of magnetic field that has undergone reconnection are red. Field lines that have not yet undergone reconnection or will not undergo reconnection are blue. The black arrow represents jet plasma that is expelled up along the reconnected open field by the interchange reconnection. Tracing the red open field line from its open end to its foot shows that the interchange reconnection transfers twist from the erupting flux-rope field to the reconnected open field in the jet. From close inspection of this cartoon, it can be seen that the erupting flux-rope field has right-handed twist and that the twist transferred to the reconnected open field is right-handed. The upper product of the interchange reconnection is the twisted open field, and the lower product is closed field added to the bipole that does not erupt. The internal tether-cutting reconnection of the legs of the erupting core field builds, unleashes, and heats the erupting flux rope above it and simultaneously builds a compact flare arcade below it. In the example blowout jet in Figure 2, we take the compact arch that brightens in the eastern foot of the jet base to be this compact flare arcade, made by internal tether-cutting reconnection in the wake of the erupting filament.

The lower-left cartoon shows the jet right after the blowing-out flux-rope field has been entirely opened by the interchange reconnection and the interchange reconnection has ended. The reconnected open field has greatly reduced the pitch of its twist by untwisting, by propagating its twist outward along itself in the form of a magnetic-untwisting wave. Because the twist in the reconnected open field is right-handed, the magnetic-untwisting wave, viewed from above, gives the field and plasma in it left-handed (clockwise) spin about the axis of the jet, as is indicated by the black swirl arrow. The long black arrow represents the jet-outflow plasma entwined in the untwisting open field. The observations of our jets lead us to think that the magnetic-untwisting wave somehow results in the plasma that it propagates through being propelled outward along the untwisting open field**,** perhaps as modeled by Shibata & Uchida (1986).

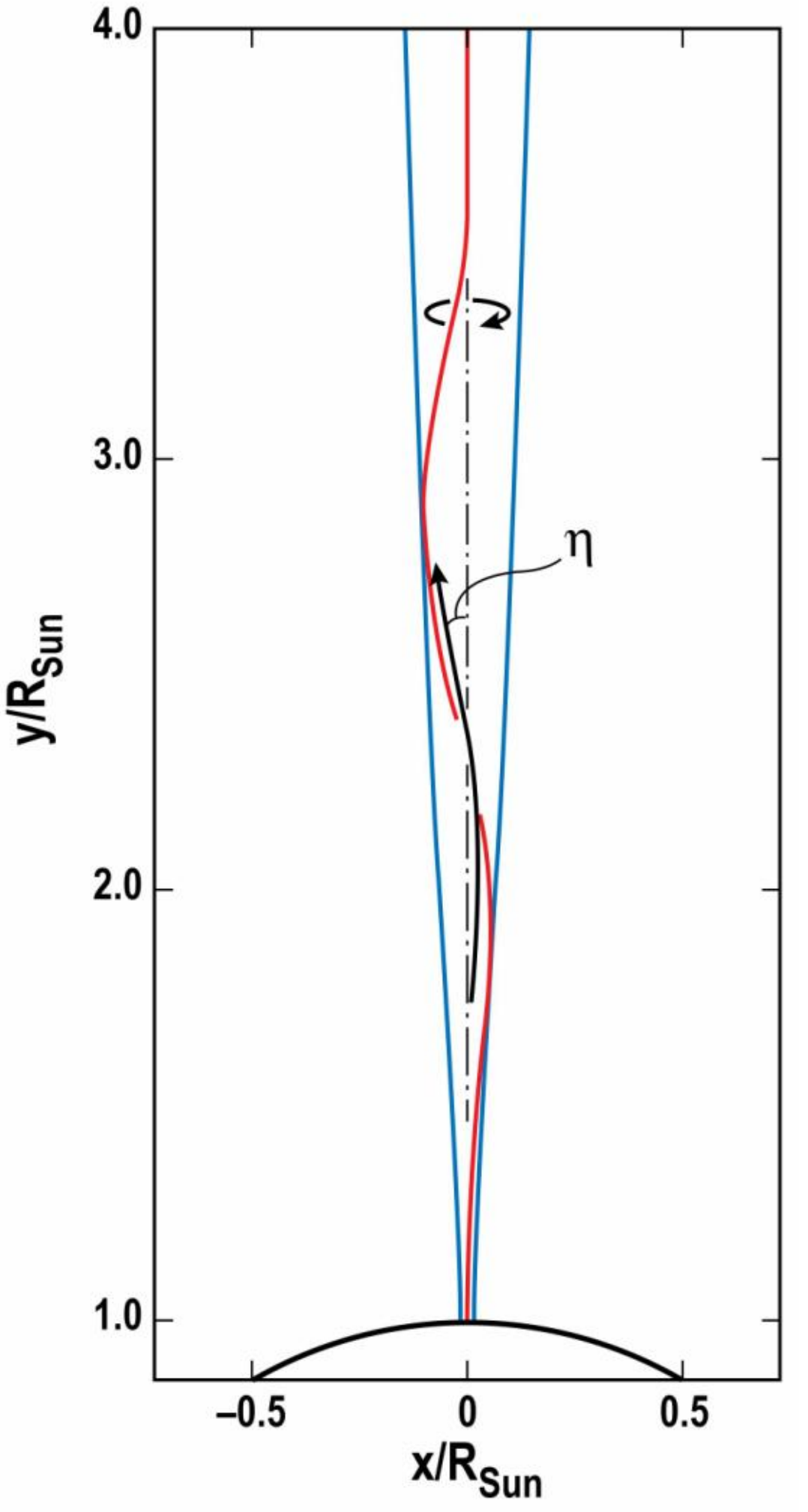

**Figure 14.** Schematic of our interpretation of the observed swaying of a C2 white-light jet as being due to the guiding of the jet outflow by the helical magnetic field in the magnetic-untwisting wave. This depiction is for when the wave and jet outflow have both reached well beyond 2 $R_{sun}$, well into the field of view of LASCO/C2.

The lower-right cartoon depicts the jet after discernible axial rotation of the jet in its AIA He II 304 Å movie has ended. By this stage, the spire of the jet seen in X-ray and EUV images has no discernible twist, and has reached or nearly reached its greatest brightness and extent or has begun to decay. By late in this stage in our 14 jets, the magnetic untwisting wave and the jet outflow have entered the outer corona viewed by LASCO/C2.

Figure 14 is our schematic of the magnetic-untwisting wave and jet outflow when they have entered the LASCO/C2 field of view beyond the edge of the C2 occulting-disk mask at 2.2 $R_{sun}$. We think that at this stage the jet plasma might continue to be driven outward by the magnetic-untwisting wave propagating through it. This cartoon depicts the special case in which the base of the jet is in the center of the northern polar coronal hole, so that the direction of the guide field, in the absence of the wave, is radial. As in Figure 13, the untwisting reconnected open magnetic field is red, the ambient open field is blue, the clockwise spin of the magnetic-untwisting wave is indicated by the black swirl arrow, and the long black arrow entwined in the untwisting field represents the jet of out-flowing plasma. The ambient-field blue lines encase the flux tube in which the magnetic-untwisting wave propagates. In this drawing, the ambient field spreads apart faster than radial out to about 3 $R_{sun}$. The angle $\eta$ is the amplitude of the angular deviation (observed in the plane of the sky) of the white-light jet's direction from the unperturbed guide field's direction, which in this cartoon is the radial direction. We assume that the jet outflow is along the helical magnetic field in the magnetic-untwisting wave, as is depicted here. The angle $\eta$ of the jet outflow then shows the pitch angle of the twist in the magnetic field in the wave, which sets the amplitude of the wave. For this picture of the magnetic-untwisting wave, the sway-sweep angle $\Delta\theta$ that we have measured from the (2.2 $R_{sun}$ to 3.0 $R_{sun}$) bottom segment of each of our jets is approximately twice $\eta$ in this distance range; in particular, at 2.3 $R_{sun}$, $\eta \approx \Delta\theta/2$.

*6.2.2. Wave Rotation Period at 2.3 $R_{Sun}$*

Our working hypothesis is that the spin observed in coronal jets is that of a magnetic-untwisting wave in the reconnected open field of the jet, and that this wave is essentially a torsional Alfven wave of circular cross section. In each of our two example jets, we can test this hypothesis by comparing the period of the observed oscillatory swaying at 2.3 $R_{Sun}$ with the rotation period of the Alfvenic magnetic-unwinding wave at 2.3 $R_{Sun}$ that has both the observed diameter of the jet and the observed amplitude of the swaying of the jet.

For a jet and torsional wave of circular cross section, the rotation period $\tau_{rot}$, the time for the magnetic-unwinding wave to undergo one 360° turn, is given by $\tau_{rot} = \pi D/V_{rot}$, where D is the diameter of the jet and wave and $V_{rot}$ is the rotation speed at the periphery of the jet and wave. We assume that the magnetic-untwisting wave approximates a torsional Alfven wave in that the energy carried in the wave is approximately half kinetic energy and half magnetic energy: $(1/2)\rho(V_{rot})^2 \approx (B_\perp)^2/8\pi$, which gives $V_{rot} \approx B_\perp/(4\pi\rho)^{1/2}$, where $\rho$ is the plasma mass density and $B_\perp$ is the twist component of the magnetic field at the periphery of the jet and wave, the component orthogonal to the direction of the unperturbed ambient magnetic field. This energy-partition condition gives $\tau_{rot} \approx 2\pi^{3/2}\rho^{1/2}D/B_\perp$ for the rotation period at any given location along the twisted flux tube of the jet and torsional wave.

As we depict in Figure 14, we assume that the angle $\eta$, the angular amplitude of the jet's swaying relative to the direction of the unperturbed guide field, is approximately the pitch angle of the torsional wave's twisted magnetic field at the periphery of the wave. That is, we assume that $\sin\eta \approx B_\perp/B$, where B is the strength of the vector magnetic field in the twisted flux tube of the wave and jet. So, for the rotation period of the wave at 2.3 $R_{Sun}$ we have $(\tau_{rot})_2 \approx 2\pi^{3/2}(\rho^{1/2})_2 D_2/(B_2 \sin\eta_2)$, where the subscript 2 denotes the value of the quantity at 2.3 $R_{Sun}$.

The flux of the magnetic field crossing an orthogonal cross section of the flux tube of the jet and wave is the product of B and the area A of the cross section. By conservation of magnetic flux in a flux tube, $B_2A_2 = B_1A_1$, where the subscript 1 denotes the value of the quantity at 1.03 $R_{Sun}$ (≈ 20,000 km above the photosphere). From $A = (\pi/4)D^2$, we have $B_2 = (D_1/D_2)^2 B_1$, and $(\tau_{rot})_2 \approx 2\pi^{3/2}(\rho^{1/2})_2(D_2)^3/[(D_1)^2 B_1 \sin\eta_2]$.

For fully-ionized coronal plasma, $\rho \approx 2 \times 10^{-24} n_e$, where $n_e$ is the electron number density. Assuming that at 2.3 $R_{Sun}$ the plasma density in the jet is of the order of the ambient plasma density, we take $(n_e)_2 \sim 2 \times 10^5$ cm$^{-3}$ (Allen 1973). As we pointed out in Section 6.2.1, we assume that $\eta_2$ is approximately half the observed sway sweep angle $\Delta\theta$: $\eta_2 \approx \Delta\theta/2$. At 1.03 $R_{sun}$ in a polar coronal hole, $B_1 \approx 10$ G (Ito et al 2010). With these approximations, we obtain our estimate of the rotation period of the magnetic-untwisting wave at 2.3 $R_{Sun}$ in terms of each jet's measured quantities, $D_1$, $D_2$, and $\Delta\theta$:

$$(\tau_{rot})_2 \approx 1.3 \times 10^{-10}\pi^{3/2}(D_2)^3/[(D_1)^2 \sin(\Delta\theta/2)] \text{ s.}$$

For our first example jet (Figures 1, 2, 3) (our second jet of 2010 August 11), from Tables 3 and 4, $D_1 \approx$ 14,000 km, $D_2 \approx$ 150,000 km, and $\Delta\theta \approx 13°$, giving $(\tau_{rot})_2 \approx$ 11,000 s ≈ 180 minutes. This estimated expected rotation period at 2.3 $R_{Sun}$ is only a factor of 2 longer than the jet's observed sway period of about 90 minutes at 2.3 $R_{Sun}$ (Figure 11, left plot). Considering the many approximations and assumptions that we employed in deriving our formula for $(\tau_{rot})_2$, we take this to be satisfactory agreement between the observed and estimated $(\tau_{rot})_2$. For our second example jet (Figures 4, 5, 6) (our jet of 2011 April 9), $D_1 \approx 22{,}000$ km, $D_2 \approx 120{,}000$ km, and $\Delta\theta \approx 8.5°$, giving $(\tau_{rot})_2 \approx 3500$ s ≈ 58 minutes, in good agreement with this jet's observed sway period of about 60 minutes at 2.3 $R_{Sun}$ (Figure 11, right plot). For only one other of our jets (the jet of 2011 June 26) was the jet detectable in enough consecutive C2 running-difference images (4 or more), and was the swaying amplitude large enough, that the swaying was discernibly oscillatory and the oscillation period could be estimated with less than about ± 50% uncertainty. The swaying of this jet appeared to be oscillatory with a period of roughly 50 minutes. For this jet, $D_1 \approx 22{,}000$ km, $D_2 \approx 120{,}000$ km, and $\Delta\theta \approx 5°$, giving $(\tau_{rot})_2 \approx 6000$ s ≈ 100 minutes, which we consider to be satisfactory agreement with the observed period.

The above agreement of the calculated $(\tau_{rot})_2$ with the oscillatory period of the observed swaying in all three of our jets in which the period could be estimated from the observed swaying bolsters our confidence in our assumptions (1) that the magnetic-untwisting waves in high-reaching solar jets such as ours are basically torsional Alfven waves, and (2) that the sway sweep shows the amplitude of the wave at 2.3 $R_{Sun}$.

### *6.2.3. Wave Energy Loss in the Inner Corona*

In this section we estimate from the observations of our 14 high-reaching jets the energy in the magnetic-untwisting wave at the bottom of the inner corona and at the top. Specifically, for any given jet, we estimate $E_1$, the wave energy per 360° turn of the wave at 1.03 $R_{Sun}$, and $E_2$, the wave energy per turn at 2.3 $R_{Sun}$. In this section, as in Section 6.2.2, the subscript 1 denotes quantities at 1.03 $R_{Sun}$, and the subscript 2 denotes quantities at 2.3 $R_{Sun}$. These two estimated energies together show whether the wave somehow loses a substantial amount of its energy in propagating up through the inner corona: the ratio $E_2/E_1$ is approximately the fraction of the wave's initial energy not lost in the inner corona, the fraction that the wave carries into the outer corona.

The wave energy E crossing an orthogonal cross section of the wave/jet flux tube during one rotation of untwisting is the product of the area A of the cross section, the wave's energy flux F per unit area, and the rotation period $\tau_{rot}$: $E = AF\tau_{rot}$. We assume that the wave energy flux is approximately that of a torsional Alfven wave, $F \approx (1/2)\rho(V_{rot})^2V_A = (4\pi^{1/2})^{-1}\rho^{1/2}(V_{rot})^2B$, which with $A = (\pi/4)D^2$ and $\tau_{rot} = \pi D/V_{rot}$, gives E in terms of $V_{rot}$, D, B, and $\rho$:

$$E \approx (\pi^{3/2}/16)\rho^{1/2}BD^3V_{rot}.$$

For equipartion of the wave's magnetic energy and kinetic energy, as in an Alfven wave, $(1/2)\rho(V_{rot})^2 \approx (B_\perp)^2/8\pi$, which with $B_\perp = B\sin\eta$ gives $V_{rot} \approx B\sin\eta\,(4\pi\rho)^{-1/2}$ and expresses E in terms of $\sin\eta$, D, and B:

$$E \approx (\pi/32)B^2D^3\sin\eta.$$

| **Table 5** Estimated Energy per Turn of Magnetic-Untwisting Wave at 1.03 $R_{Sun}$ and at 2.3 $R_{Sun}$ | | | |
|---|---|---|---|
| Event | $E_1$ (ergs) | $E_2$ (ergs) | Energy Ratio $E_2/E_1$ |
| Example jet of 2010 August 11 | (1.7, **3.3**, 5.9) x $10^{27}$ | (1.2, **2.8**, 6.2) x $10^{26}$ | 0.02, **0.09**, 0.36 |
| Example jet of 2011 April 9 | (1.2, **1.8**, 2.7) x $10^{28}$ | (0.74, **1.4**, 2.6) x $10^{27}$ | 0.03, **0.08**, 0.22 |
| Average jet | (5.3, **9.1**, 12) x $10^{27}$ | (1.9, **4.5**, 9.4) x $10^{26}$ | 0.02, **0.05**, 0.18 |

Note: $E_1$ is the wave's energy per turn at 1.03 $R_{Sun}$, and $E_2$ is the energy per turn at 2.3 $R_{Sun}$.

We use the expression of E in terms of $V_{rot}$, D, B, and $\rho$ to estimate $E_1$. At 1.03 $R_{Sun}$ in big polar coronal jets such as ours, $B_1 \approx 10$ G (Ito et al 2010), $\rho_1 \approx 2.0 \times 10^{-24} (n_e)_1$ gm cm$^{-3}$, and $(n_e)_1 \approx 6 \times 10^8$ cm$^{-3}$ (Pucci et al 2013), from which we obtain:

$$E_1 \approx 1.2 \times 10^{-7} (D_1)^3 (V_{rot})_1 \text{ ergs.}$$

We use the expression of E in terms of $\sin\eta$, D, and B to estimate $E_2$. By conservation of magnetic flux in the wave/jet flux tube, $B_2 = B_1(D_1/D_2)^2$, from which, with $B_1 \approx 10$ G and $\eta_2 \approx \Delta\theta/2$, we obtain:

$$E_2 \approx 9.8 (D_1)^4[\sin(\Delta\theta/2)]/D_2 \text{ ergs.}$$

From the values of $(V_{rot})_1$ and $D_1$ and their estimated uncertainties given in Table 3 and the values of $D_2$ and $\Delta\theta$ and their estimated uncertainties given in Table 4, we obtain from the above formulas for $E_1$ and $E_2$ the estimated per-turn wave energies ($E_1$ and $E_2$) and their ratio ($E_1/E_2$) given in Table 5 for each of our two example jets and for the average jet, the hypothetical jet having the average values of $(V_{rot})_1$, $D_1$, $D_2$, and $\Delta\theta$ of our 14 high-reaching jets. For the average jet, we take the uncertainty in $(V_{rot})_1$, $D_1$, $D_2$, and $\Delta\theta$ to be ± 20 km s$^{-1}$, ± 2 x $10^3$ km, ± 10 x $10^3$ km, and ± 1.5°, respectively. These are typical values for our 14 jets (see Tables 3 and 4). In Table 5, three values are given in each entry: the middle value (in bold face) is the nominal value obtained from the values of $(V_{rot})_1$, $D_1$, $D_2$, and $\Delta\theta$ listed for that jet in Tables 3 and 4; the smaller value (on the left) is the smallest value permitted by the estimated uncertainties in $(V_{rot})_1$, $D_1$, $D_2$, and $\Delta\theta$ for that jet; the larger value (on the right) is the largest value permitted by the estimated uncertainties in $(V_{rot})_1$, $D_1$, $D_2$, and $\Delta\theta$ for that jet. As Table 5 shows, in each of the three cases, we estimate that only of order 10% of the energy per turn at 1.03 $R_{Sun}$ in the bottom of the inner corona remains in the wave at 2.3 $R_{Sun}$, which is at the top of the inner corona, that is, near the bottom of the outer corona viewed by LASCO/C2. Thus, by interpreting the observed spinning in the eruption of our jets and the observed swaying of our jets in the LASCO/C2 outer corona as being manifestations of an Alfvenic magnetic-untwisting wave, we find that these waves lose most of their energy in the inner corona below 2.3 $R_{Sun}$.

## 7. SUMMARY AND DISSCUSSION

From the observations of large EUV/X-ray polar coronal jets presented and reviewed in this paper we propose that:

1. The jet outflow is entwined in a co-produced torsional magnetic wave that makes the erupting jet spin.
2. Interchange reconnection of twisted closed magnetic field in the base of the jet with impacted ambient open field transfers twist from the closed field to the reconnected open field in the jet.
3. The reconnected open field then untwists by propagating its twist outward along itself as a torsional magnetic wave.

Also from the jet observations presented and reviewed in this paper we infer that:

1. All but a few jets that erupt in coronal holes either do not reach the outer corona beyond 2.2 $R_{Sun}$ or are too faint there to be visible in LASCO/C2 running-difference images. The few jets that are visible in LASCO/C2 running-difference images have exceptionally large amounts of untwisting. This suggests that the magnetic-untwisting wave somehow drives the jet plasma outflow.
2. In most polar jets that are visible in LASCO/C2 running-difference images, the jet outflow has measurable swaying in sequences of these images. The amplitude of the swaying is plausibly the amplitude of the untwisting wave in the magnetic field that guides the jet outflow in the LASCO/C2 outer corona.

From analysis of *Hinode*/XRT movies of large X-ray jets in polar coronal holes, Cirtain et al (2007) measured jet-front speeds of ~ 800 km $s^{-1}$ and oscillatory swaying of the jet having periods of ~ 200 s and swaying speeds of ~ 40 km $s^{-1}$. They interpreted these properties to be evidence of Alfven waves in these jets. Our 14 large polar jets are broadly consistent with this interpretation and add some refinements. We have found evidence that each of our jets has a magnetic wave in it, and that the wave is a torsional wave, a magnetic-untwisting wave. The observed spin of our erupting jets and the observed swaying of the jets in the outer corona together imply that the wave is grossly a torsional Alfven wave and that the wave loses most of its energy in the inner corona below 2.3 $R_{Sun}$. This and the supersonic speeds of the jet fronts in transiting the inner corona suggest that the wave drives the jet outflow, thereby producing a jet-front speed that can approach the Alfven speed in the jet-front plasma.

Provided that the magnetic-untwisting wave in our jets is grossly a torsional Alfven wave in the manner we have assumed, we have found, from the observed rotation speed and diameter of the jets at the bottom of the corona and the observed sway sweep and jet diameter at the top of the inner corona (at 2.3 $R_{Sun}$), that the wave loses most of its energy in the inner corona (Table 5). From this we infer that as the wave propagates up through the inner corona it gives most of its energy, by some process, to the plasma through which it propagates. We expect that the energy-transfer process entails non-linear aspects of the wave. One candidate process is intermittent magnetic levitation of the coronal plasma by the magnetic-untwisting waves: the wave in each jet lifts the plasma that it propagates through (drives the jet outflow) and after the passage of the wave the elevated rising plasma is slowed and stopped by gravity and then falls back and heats the corona by in-fall impact (e.g., Hollweg et al 1982; Moore et al 1992). [MHD modeling has indicated that a non-linear magnetic-untwisting wave in a vertical flux tube in the low corona can propel the plasma upward at speeds of order the Alfven speed (e.g., Shibata &

Uchida 1986). See Moschou et al (2013) for evidence of an upward force in large jets observed in AIA He II 304 Å movies.] Another candidate for the energy-transfer process is turbulent cascade of the wave to Alvenic waves of progressively higher wavenumbers, finally dissipating at the high-wavenumber end of the cascade to heat the plasma in the cascading turbulence (e.g., Hollweg 1984; van Ballegooijen et al 2011). We expect that physically realistic numerical modeling of the generation and non-linear propagation of the magnetic-untwisting waves in polar coronal jets will be required to establish how these waves lose energy in the inner corona.

It is observed that Type-II spicules have axial-rotation spin with rotation speeds of 25-30 km $s^{-1}$ (De Pontieu et al 2012). From analysis of the proper motions in Type-II spicules at the limb in a polar coronal hole in a *Hinode*/SOT Ca II H movie, Tavabi et al (2015) have recently presented evidence that the spin is from untwisting of the magnetic field like the untwisting we observe in our big jets. Moore et al (2011) proposed that Type-II spicules are miniature versions of the large EUV/X-ray jets in coronal holes, made in the same way by interchange reconnection of emerging and blowout-erupting granule-size closed magnetic bipoles (~ 1000 km in diameter) with the ambient feet of the coronal magnetic field rooted in the magnetic network in quiet regions and corona holes. From the observed number of granule-size bipoles and the observed plasma density and magnetic field strength in the chromosphere in polar coronal holes, Moore et al (2011) estimated that (if the spin in Type-II spicules is due to magnetic-unwinding waves that are basically torsional Alfven waves) the Alfvenic waves from Type-II spicules, on average over areas larger than a network cell (~30,000 km in diameter), carry an energy flux of ~ $3 \times 10^5$ erg $cm^{-2}$ $s^{-1}$ into the corona in polar coronal holes. This is of the order of the mechanical energy flux needed to power the coronal heating in coronal holes (Withbroe 1988). Thus, the corona in coronal holes can be heated by the torsional magnetic waves from Type-II spicules provided that these waves dissipate in the inner corona instead of just passing through to the solar wind. If the torsional waves in Type-II spicules are made the same way as in our big polar jets, then they plausibly lose energy in the inner corona by the same process as in our big jets. Thus, our finding that the magnetic-untwisting waves in our big jets lose most of their energy in the inner corona raises the posssibility that the magnetic-untwisting waves from Type-II spicules likewise dissipate in the inner corona and thereby power most of the coronal heating in coronal holes.

Our inference that Type-II spicules plausibly power the coronal heating in coronal holes via their torsional magnetic waves is supported by recent analysis by Hahn & Savin (2013) of coronal EUV spectral line widths as a function of height in a polar coronal hole. From the line widths observed by the Extreme ultraviolet Imaging Spectrometer (EIS, Culhane et al 2007) on *Hinode*, they estimate that the area-average energy flux of Alfven waves entering the base of the corona in the polar coronal hole is about $7 \times 10^5$ erg $cm^{-2}$ $s^{-1}$, and find that the waves decay with height, losing about 85 % of their energy below 1.5 $R_{Sun}$ from Sun center.

This work was funded by the Heliophysics Division of NASA's Science Mission Directorate through the Living With a Star Targeted Research and Technology Program. ACS benefited from discussions held at the International Space Science Institute's (ISSI; Bern, Switzerland) International Team on Solar Coronal Jets.